\documentclass[iop,apj]{emulateapj}

\begin{document}

\newcommand{\kms}{\ensuremath{\mathrm{km}\,\mathrm{s}^{-1}}}
\newcommand{\MLsun}{\ensuremath{\mathrm{M}_{\sun}/\mathrm{L}_{\sun}}}
\newcommand{\Lsun}{\ensuremath{\mathrm{L}_{\sun}}}
\newcommand{\Msun}{\ensuremath{\mathrm{M}_{\sun}}}
\newcommand{\Aunits}{\ensuremath{\mathrm{M}_{\sun}\,\mathrm{km}^{-4}\,\mathrm{s}^{4}}}
\newcommand{\gevcc}{\ensuremath{\mathrm{GeV}\,\mathrm{cm}^{-3}}}
\newcommand{\etal}{et al.}
\newcommand{\LCDM}{$\Lambda$CDM}
\newcommand{\ML}{\ensuremath{\Upsilon_*}}

\shorttitle{Weighing Galaxy Disks}
\shortauthors{McGaugh \& Schombert}

\title{Weighing Galaxy Disks with the Baryonic Tully-Fisher Relation}

\author{Stacy S. McGaugh}
\affil{Department of Astronomy, Case Western Reserve University, Cleveland, OH 44106}
\email{stacy.mcgaugh@case.edu} 

\and

\author{James M. Schombert}
\affil{Department of Physics, University of Oregon, Eugene, OR 97403}
\email{jschombe@uoregon.edu}

\begin{abstract}
We estimate the stellar masses of disk galaxies with two independent methods:
a photometrically self-consistent color--mass-to-light ratio relation (CMLR) from population synthesis models, 
and the Baryonic Tully-Fisher relation (BTFR) calibrated by gas rich galaxies. 
These two methods give consistent results.  The CMLR correctly converts distinct 
Tully-Fisher relations in different bands into the same BTFR.
The BTFR is consistent with $M_b \propto V_f^4$ over nearly six decades in mass, with no hint of a change in slope over that range.
The intrinsic scatter in the BTFR is negligible, implying that the IMF of disk galaxies is effectively universal. 
The gas rich BTFR suggests an absolute calibration of the stellar mass scale that yields nearly constant mass-to-light ratios in the near-infrared (NIR): 
$0.57\;\MLsun$ in $K_s$ and $0.45\;\MLsun$ at $3.6\micron$. There is only modest intrinsic scatter ($\sim 0.12$ dex) about these typical values.
There is no discernible variation with color or other properties: the NIR luminosity is a good tracer of stellar mass.  
\end{abstract}

\keywords{galaxies: evolution --- galaxies: fundamental parameters --- galaxies: photometry --- galaxies: kinematics and dynamics --- galaxies: stellar content}

\section{Introduction}
\label{sec:intro}

The stellar mass of a galaxy is one of its most fundamental characteristics.  
Based on our knowledge of stellar evolution, we expect to be able to use measured galaxy luminosity and color to estimate stellar mass
\citep[e.g.,][]{BdJ01,Bell03,Port04,Zib09,IP13}.
However, population synthesis models remain uncertain \citep{conroy}, and are not always self-consistent \citep{MS2014}.

Despite considerable progress, a variety of difficult issues persist.  
The star formation history of unresolved populations is difficult to uniquely constrain.
The contribution to the integrated luminosity of bright but short-lived stars 
in late stages of evolution remains uncertain \citep[e.g., AGB stars:][]{maraston05,Marigo2008,Kriek10,Melbourne12,Zib13}.
The distribution of stellar metallicity as well as the mean value appears to play an 
important role \citep{rakosschombert09a,conroy,SM14pop,SM2014}.
Stellar mass estimates can depend considerably on the treatment of these effects.

The largest persistent uncertainty is in the mass spectrum of stars (the IMF).  
Most of the light is produced by massive stars while most of the integrated mass
resides in low mass stars.  Consequently, small differences in the IMF can have large effects on the stellar masss-to-light ratio, \ML.
Such variations would lead to substantial scatter in the Tully-Fisher relation \citep{TForig}, which not observed.
This implies that the galaxy-averaged IMF does not vary wildly \citep{btforig,verhTF}.  

Methods to estimate stellar mass that are independent of the IMF would be helpful.
One approach to this problem is to measure the vertical velocity dispersion in face-on disk galaxies \citep[e.g.,][]{bershady10,bershady11}.
Another is provided by the Baryonic Tully-Fisher Relation (BTFR).
This is one of the strongest correlations in extragalactic astronomy \citep{M05}, 
providing an empirical link between mass and observed rotation velocity.

Recent work on gas dominated galaxies \citep{begum,swaterswhisp,stark,trach} has 
made it possible to obtain an absolute calibration of the mass scale of the 
BTFR that is largely independent of stellar mass estimates \citep{M12}.  When the gas mass exceeds the stellar mass, the systematic error
in the stellar mass induced by the IMF, etc., becomes a minor contributor to the error budget \citep{M11}.  
The assumptions necessary in population models cease to be the dominant factor in the calibration of the BTFR.  

Here we examine a sample of galaxies with high quality, extended rotation curves from 21cm interferometry and 
photometry in the optical and near-infrared (\S \ref{sec:data}). We first construct the luminous Tully-Fisher relation for each band,
and the corresponding stellar mass Tully-Fisher relation and BTFR using self-consistent stellar population mass-to-light 
estimators (\S \ref{sec:fits}).  We then estimate the stellar mass of each galaxy using the BTFR independently calibrated 
by gas rich galaxies (\S \ref{sec:btfrmass}), and use this as a check on the IMF (\S \ref{sec:IMF}).  
We compare these results and discuss their consistency (or lack thereof) with other
methods (\S \ref{sec:othermeth}), and summarize in \S \ref{sec:conc}. \\

\section{Data}
\label{sec:data}

The data used here are based on the sample of galaxies 
with extensive photometry in optical and near-infrared (NIR) bands assembled by \citet{MS2014}. 
All galaxies employed here have been observed at $B$, $V$, and [3.6].  
Many, though not all, also have $I$, $J$, and $K_s$-band data.
Many SINGS \citep{KSINGS,DSINGS} galaxies observed 
as part of the THINGS \citep{FTHINGS} HI survey are utilized. 
The sample further includes many new Spitzer observations \citep{SM2014}.
These data are used to estimate the luminosity and stellar mass of each galaxy.

We select galaxies from \citet{MS2014} for which credible measurements of the outer, 
flat rotation velocity $V_f$ are available from 21 cm interferometry.  This measure provides the best available estimate of the characteristic
rotation velocity of disk galaxies in the sense that it minimizes the scatter in the Tully-Fisher relation \citep{verhTF,M05}.
The more commonly employed line-width is less precise, in that it results in a larger irreducible scatter in the BTFR \citep{verhTF,M12,Zarit2014}.
This difference is critical for estimating stellar masses from the BTFR, as uncertainty in the rotation velocity propagates strongly into
the implied mass.

\begin{deluxetable*}{lccccccr}
\tablewidth{0pt}
\tablecaption{Rotation Velocities and Gas Masses}
\tablehead{
\colhead{Galaxy}  & \colhead{$V_f$} & \colhead{$M({HI})$} & \colhead{$M({H_{2}})$}
& \colhead{$L_V$} & \colhead{$L_I$} & \colhead{$L_{[3.6]}$} & \colhead{Refs.} \\
& (\kms) & \multicolumn{2}{c}{$(10^9\;\mathrm{M}_{\sun})$} & \multicolumn{3}{c}{$(10^9\;\mathrm{L}_{\sun})$} &
 }
\startdata
DDO 154  & $53.3\pm4.7$  &0.0925 &0.00618& 0.0516  &0.0294  &0.0725  &1,2,3,4,5 \\
D631-7   & $52.9\pm5.0$  &0.147  &0.0127 & 0.0540  &0.0448  &0.0975  &1,2,3,6 \\
DDO 168  & $53.4\pm2.5$  &0.317  &0.00279&  0.166  & \dots  & 0.187  &1,2,3,7 \\
D500-2   & $68.1\pm6.6$  &0.871  &0.00790&  0.304  & 0.226  & 0.395  &1,2,6  \\
NGC 2366 & $59.7\pm10.3$ &0.600  &0.123  &  0.456  & 0.369  & 0.716  &1,5,8  \\
IC 2574  & $76.7\pm4.5$  &1.41   &0.0937 &   1.04  & 0.956  &  2.33  &1,5,8  \\
F563-1   &$111.\pm10.$   &3.50   &0.0900 &   1.13  &  1.25  &  2.84  &1,2,3,7 \\
NGC 2976 &$ 86.0\pm3.5$  &0.134  &0.121  &   1.13  &  1.05  &  3.19  &1,4,5 \\
F568-V1  &$124.\pm10.$   &3.93   &0.134  &   2.36  &  2.25  &  4.20  &1,2,3,7 \\
NGC 1003 &$113.5\pm1.9$  &5.31   &0.243  &   2.76  & \dots  &  5.91  &1,4,5 \\
F568-1   &$116.\pm10.$   &4.02   &0.141  &   3.10  &  2.98  &  7.10  &1,2,3,9 \\
NGC 7793 &$110.4\pm4.4$  &0.761  &0.282  &   2.99  &  1.80  &  7.55  &1,4,5 \\
UGC 128  &$131.\pm10.$   &6.51   &0.420  &   4.49  &  4.20  &  11.3  &1,2,3,10,11 \\
NGC 2403 &$135.8\pm4.0$  &2.52   &0.522  &   3.86  &  5.00  &  12.1  &1,4,5,10,11 \\
NGC 925  &$113.6\pm6.3$  &3.99   &0.826  &   8.35  &  8.35  &  15.8  &1,4,5 \\
NGC 3198 &$148.9\pm4.2$  &0.134  &1.30   &   12.4  &  15.8  &  31.2  &1,4,5 \\
NGC 3621 &$152.3\pm3.2$  &6.98   &7.06   &   6.74  &  7.11  &  33.1  &1,4,5   \\
NGC 3521 &$191.9\pm17.6$ &4.48   &1.65   &   15.6  &  23.2  &  91.1  &1,4,5 \\
NGC 3031 &$199.0\pm13.8$ &6.97   &1.58   &   23.0  &  38.7  &  106.  &1,4,5 \\
NGC 5055 &$181.1\pm11.9$ &7.21   &2.36   &   26.3  &  40.2  &  135.  &1,4,5 \\
NGC 2998 &$211.7\pm2.7$  &27.7   &7.27   &   74.8  & \dots  &  160.  &1,7 \\
NGC 6674 &$240.5\pm4.2$  &38.2   &9.30   &   63.2  & \dots  &  230.  &1,7 \\
NGC 7331 &$245.8\pm8.4$  &9.38   &4.30   &   38.3  &  67.1  &  259.  &1,4,5  \\
NGC 801  &$218.8\pm3.5$  &36.7   &12.0   &   71.5  & \dots  &  268.  &1,7 \\
NGC 5533 &$239.9\pm5.0$  &23.5   &13.9   &   62.7  & \dots  &  303.  &1,7  \\
UGC 2885 &$298.\pm10.$   &34.5   &15.7   &   179.  & \dots  &  468.  &1,7
\enddata
\tablerefs{1.~\citet{MS2014}. 2.~\citet{SM2014}. 3.~\citet{SMM11}. 4.~\citet{THINGS}. 5.~\citet{leroy}.
6.~\citet{trach}. 7.~\citet{M05}. 8.~\citet{OhThings}. 9.~\citet{dBMH96}. 10.~\citet{dBM1996}. 11.~\citet{VdB1999}.
}
\tablecomments{Galaxies are listed in order of increasing [3.6] luminosity as in \citet{MS2014}.}
\label{basicdata}
\end{deluxetable*}

We selected galaxies for observation with Spitzer \citep{SM2014} to extend the dynamic range over which the BTFR could be constructed.
Galaxies with existing high quality HI rotation curves were selected to supplement the THINGS sample \citep{FTHINGS} at both ends of the BTFR.
The resulting sample extends to both greater and lower luminosity than present in THINGS.  The aim was to cover as much range in mass 
as possible with evenly sampled bins.  This is done to avoid the common selection effects afflicting 
magnitude limited samples, which typically have many $L^*$ galaxies but few or none with $L < 10^9\;\Lsun$.
The result is a sample of galaxies that probes the BTFR over a large dynamic range in a uniform way, free of the usual bias in which
the region around $L^*$ is probed many times while either end of the relation is under-represented.

The emphasis here is to uniformly sample as much of the BTFR as possible with the highest quality data available.  In this vein,
we require that galaxies be sufficiently inclined that $\sin(i)$ corrections are modest \citep[$i > 45^{\circ}$; see discussion in][]{stark}.  
This excludes a number of THINGS galaxies with otherwise good data (e.g., NGC 6946).
We also exclude two galaxies for which we do not entirely trust the data: NGC 2841 and NGC 2903. 
NGC 2841 is a well-known outlier from Tully-Fisher relation \citep{bottema02}.  Whether NGC 2841 is a true outlier, 
or some systematic error is to blame, is beyond the scope of this paper.  However, it is not simply at one extreme of the scatter;
it sits well away from all other data.  We therefore suspect some systematic error is to blame.
In the case of NGC 2903, the [3.6] photometry is suspect.  \citet{THINGS} note problems constructing a mass model,
and the resulting [3.6] magnitude makes NGC 2903 an outlier from both the ($V-[3.6]$, [3.6]) color--magnitude relation \citep{SM14pop} and the [3.6]
Tully-Fisher relation, even though it does not deviate from the Tully-Fisher relation in other bands, including $K_s$.  We therefore infer that the [3.6]
magnitude of NGC 2903 is not reliable, so we do not consider it further.  

The final sample contains 26 galaxies spanning the range $50 \lesssim V_f \lesssim 300\;\kms$, 
$3 \times 10^7 \lesssim M_* \lesssim 3 \times 10^{11}\;\Msun$, and $10^8 \lesssim M_g \lesssim 5 \times 10^{10}\;\Msun$.
It spans four decades in luminosity, while magnitude limited Tully-Fisher samples typically span two \citep[e.g.,][]{Pizagno07}.
This large dynamic range is more important for constraining the slope of the BTFR than is the sheer numbers of galaxies.

Sample selection is driven by data quality: it is constructed to be as free of systematic errors as possible. 
Of course, one can never completely exclude the presence of systematic errors.
The two that concern us most are residual systematic errors is in the adopted distances and in the inclination determinations.  

Direct distances measurements via Cepheids or the TRGB are adopted when available \citep{EDD}.
When no direct measurement is available, we adopt a Hubble flow estimate consistent with independent local galaxy 
data \citep[$H_0 = 75\;\kms\,\mathrm{Mpc}^{-1}$:][]{Sorce2012}.  Galaxies lacking direct measurements tend to be the most distant, 
luminous galaxies or the less studied dwarfs --- i.e., those at the extremes of the relation.  Dividing the sample into those objects
with and without direct distance determinations reveals no statistically significant difference: we recover the local distance scale.
If instead $H_0 = 67\;\kms\,\mathrm{Mpc}^{-1}$, the primary systematic effect would be to increase our mass estimates 
by 0.1 dex with a corresponding translation of the BTFR zero point.

The effects of systematic inclination errors are discussed at length by \citet{M12}.  
These most commonly manifest as outliers to the low velocity side of the BTFR.  
Oval distortions can make galaxies look less face-on than they are, so the inclination correction may be underestimated.
This is mitigated here by the use of velocity fields, which give an additional constraint on the inclination.
As we will see, the scatter around the BTFR relation is small with no obvious asymmetry, 
so systematic inclination errors appear to be small in this sample.  

Fundamental data like distances, magnitudes, and colors are given in \citet{MS2014}.  
Data specific to our present analysis are given in Table~\ref{basicdata}.
Column 1 gives the name of the galaxy.  Column 2 gives the rotation velocity $V_f$.
Columns 3 and 4 give the atomic and molecular gas masses, respecively.
Luminosities in each of the $V$, $I$, and [3.6] bands are given in columns 5 -- 7.
As in \citet{MS2014}, we adopt solar absolute magnitudes of 4.83, 4.08, and 3.24 in the $V$, $I$, and [3.6] bands.
References to the sources of the data are given in column 8.

\begin{figure*}
\epsscale{1.0}
\plotone{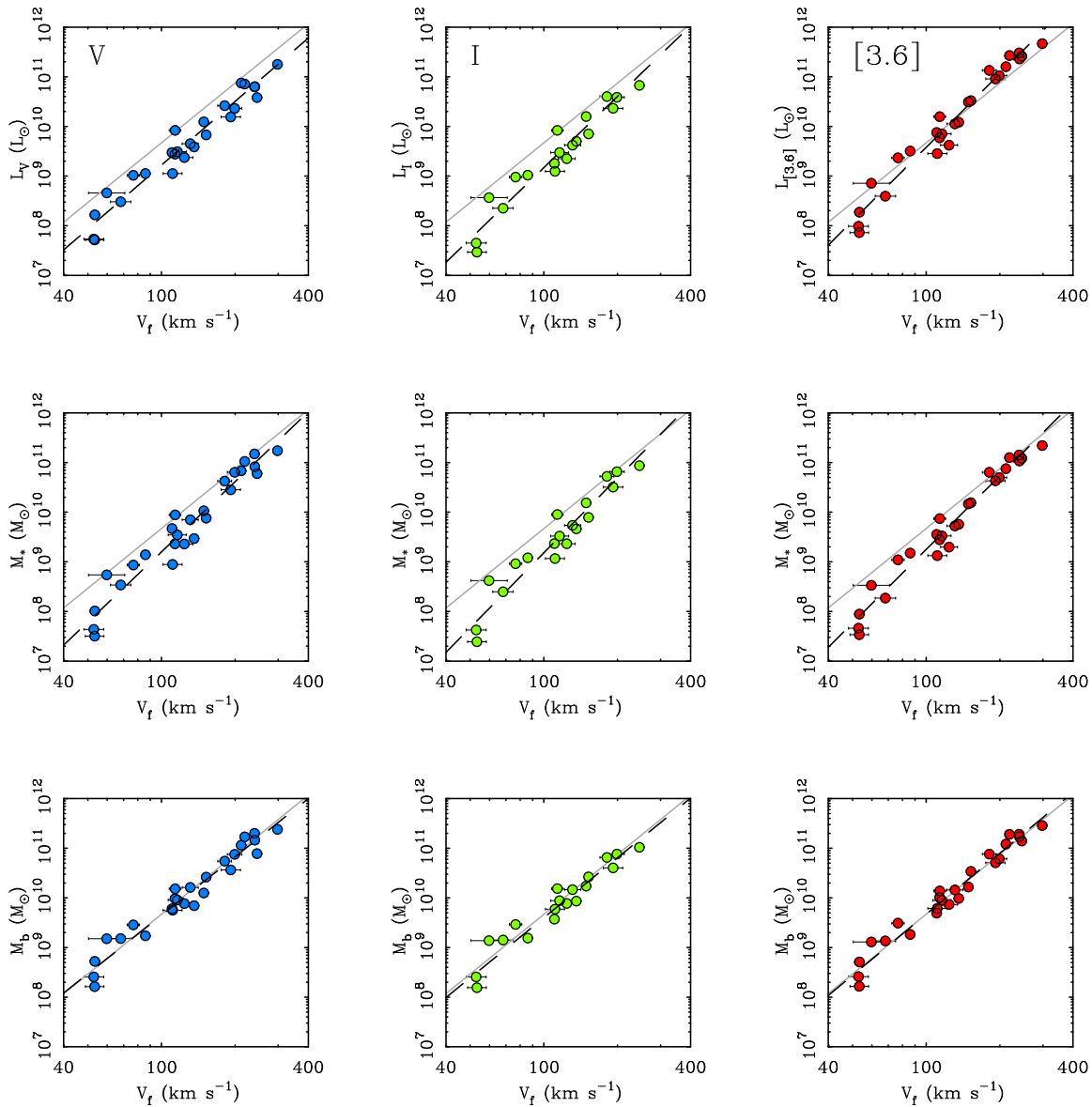}
\caption{Tully-Fisher relations determined from $V$ (left), $I$ (middle), and [3.6] (right) data for the galaxies in Table~\ref{basicdata}.
The top row shows the luminous Tully-Fisher relation (LTFR), the middle row the stellar mass Tully-Fisher relation (STFR), 
and the bottom row the baryonic Tully-Fisher relation (BTFR).
In all cases the luminosity or mass is plotted against the flat, outer velocity.  Fits to each relation are shown as dashed lines and reported
in Table~\ref{Fitstats}.  For reference, the BTFR previously fit to gas rich galaxies by \citet{M12} is shown as a light solid line in all panels.  
\label{TF}}
\end{figure*}

Rotation curves generally experience an extended radial range over which the rotation velocity is approximately constant.
This flat velocity, $V_f$, is an obvious measure of the characteristic global rotation speed of a galaxy.
We have measured or remeasured anew $V_f$ for all galaxies considered here by taking
the mean of data along the outer, flat portion of the rotation curve.  
While the outer parts of rotation curves are very nearly flat, they are not perfectly so.
We implemented several algorithms for deciding what constitutes the flat portion of the rotation curve:
a subjective estimate by eye, a definition by slope such that $\partial \log V/\partial \log R < 0.1$ \citep{M05,swaterswhisp,stark}, 
and an outlier rejection routine 
that iteratively excludes points from the rising or falling portion of the rotation curve.  These methods yield consistent results within the errors.  
Indeed, the relative variation around $V_f$ along a rotation curve, while clearly perceptible, is typically smaller than the absolute
uncertainty in $V_f$: $\Delta V = \Delta R (\partial V/\partial R) < \sigma_V$.  
Hence the mildly non-zero slopes $\partial V/\partial R$ that are observed do not preclude accurate measurement of $V_f$.
A greater concern is the occasional warp that causes a radial variation in the inclination within a galaxy.
Indeed, this might be one issue in NGC 2841.  The galaxy for which this concern is most important in the present sample is
DDO 154 \citep[see discussion in][]{THINGS}.  

The gas mass of each galaxy is adopted from the corresponding tabulated reference in Table~\ref{basicdata}.  
The HI mass is obtained from the 21 cm flux in the usual fashion: $M(HI) = 2.36 \times 10^5 D^2 F_{HI}$.
The molecular gas mass has been estimated for the THINGS galaxies by \citet{leroy}, 
who discuss in detail the conversion from observed CO flux to H$_2$ mass.  

Many galaxies lack CO detections.  In these cases, we exploit the near-constant efficiency of star formation observed by
\citet{leroy} to estimate the molecular gas mass from the observed star formation rate \citep{SMM11}.  
This method implicitly corrects for the metallicity dependence of the CO-to-H$_2$ conversion factor,
albeit by assuming a constant star formation efficiency.
For galaxies lacking CO detections, we estimate the molecular gas mass as
\begin{equation}
\log[M(H_2)] = \log(\mathrm{SFR}) + 9.15.
\end{equation}
Here, the molecular gas mass is in solar masses and the star formation rate SFR is in $\Msun\;\mathrm{yr}^{-1}$.
These estimates are rather uncertain \citep[compare to other estimators discussed by][]{M12}, 
but the molecular gas mass is usually a small portion of the mass budget, and makes little difference to the result.

The luminosities given in Table~\ref{basicdata} are computed directly from the distances, magnitudes, and colors given in \citet{MS2014}.
The [3.6] luminosities originate either with \citet{THINGS} or \citet{SM2014}, with other bands as cited in the table.
The formal uncertainties in the measured magnitudes are small, of order 0.02 mag.  The dominant uncertainty in the luminosity
and the gas mass is the distance.  We treat the uncertainties in the luminosities as negligibly small, and consider the 
random distance uncertainties as a contributor to the scatter in the Tully-Fisher relations we construct.

\section{Direct Fits}
\label{sec:fits}

\subsection{Types of Tully-Fisher Relations}
\label{sec:TTFR}

The term ``Tully-Fisher Relation'' has been used to mean a variety of subtly different things.
Most commonly, the Tully-Fisher Relation is posed as a relation between absolute magnitude and line-width \citep{TForig}
and employed as a distance indicator \citep[e.g.,][]{Sorce2012}.  Here we construct several 
distinct flavors of Tully-Fisher relations from the assembled data in hopes of identifying the underlying physical basis
of the relation.

We start with the relation between luminosity and asymptotic rotation velocity in each band.  
This is analogous to the common Tully-Fisher relation, but differs from it in that $V_f$
from extended, resolved rotation curves is used rather than line-widths.  
While similar, these are not identical measures of the characteristic rotation velocity of a galaxy.
The primary consequence is a systematic difference in the slopes of the fitted relation.
Line-widths are sensitive to the presence of inner humps in rotation curves, which are prominent in luminous galaxies
but largely absent in faint ones \citep[see, e.g,][]{verhTF,noordTF}.  This causes the $L$-$W$ relation to be shallower
than the $L$-$V_f$ relation, which we will refer to as the LTFR.

Next, we construct the stellar mass Tully-Fisher relation (STFR).
We employ stellar population models to estimate the stellar mass of each galaxy based on their observed
luminosities and the color--mas-to-light ratio relation (CMLR) of \citet{MS2014}.  We then construct the STFR implied independently by the luminosity
observed in each band pass.  If the our semi-empirical CMLR are correct, then each band's LTFR should transform into the same STFR.

Finally, we construct the baryonic Tully-Fisher Relation (BTFR).
This is a relation between rotation velocity and baryonic mass \citep{btforig,verhTF,BdJ01,pfennBTF,M05},
where now the mass includes all observed forms of baryonic mass, stars and gas: $M_b = M_*+M_g$.
Again, if all is well, the same relation should be obtained independently from each band.
We can also compare the result to that obtained independently from fitting gas rich galaxies, 
which is very nearly independent of the stellar mass estimate \citep{M11,M12}.
The LTFR, STFR, and BTFR are shown in Fig.~\ref{TF}.

The use of $V_f$ as the velocity measure rather than line-widths leads to systematically steeper slopes for all flavors of the relation.
For the BTFR, \citet{M12} found a slope $x = 3.41\pm0.08$ when line-widths are employed as the velocity measure, 
consistent with the $3.5\pm0.2$ reported independently by \citet{Zarit2014}.  
In contrast, the slope of the BTFR is very close to 4 when $V_f$ is employed as the measure of rotation velocity \citep{stark}.
It also consistently yields relations with lower scatter.  For example,
\citet{M12} estimates an irreducible scatter of $\sim 0.2$ dex in mass in the BTFR when line-widths are employed,
but negligible intrinsic scatter with $V_f$.

\subsection{Fitting}
\label{sec:fit}

For the various possible ordinates of the Tully-Fisher relation, we fit relations of the form
\begin{equation}
\log(\ell) = \log A+ x \log(V_f).
\label{eqn:BTFR}
\end{equation}
Here $A$ is the normalization of the relation ($\log A$ being the intercept in the log-log plane) and $x$ is the slope.
The symbol $\ell$ represent any of the luminosity, stellar mass, gas mass, or baryonic mass, while 
$V_f$ is of course the rotation velocity measured in the outer portion of the resolved rotation curve.

\begin{deluxetable}{ccccc}
\tablewidth{0pt}
\tablecaption{Tully-Fisher Relations and their Scatter}
\tablehead{
\colhead{Band} & \colhead{$\ell$} & \colhead{$x$} & \colhead{$\log A$} & \colhead{$\sigma_{\log \ell}$} 
}
\startdata
\multicolumn{5}{l}{\underline{Luminous Tully-Fisher Relation}} \\
$V$ & $L_V$ & $4.27\pm0.19$ & $\phm{-}0.68\pm0.41$ & 0.18  \\
$I$ & $L_I$ & $4.77\pm0.30$ & $-0.37\pm0.62$ & 0.19   \\
3.6 & $L_{[3.6]}$ & $4.93\pm0.17$ & $-0.28\pm0.36$ & 0.15  \\
\multicolumn{5}{l}{~} \\
\multicolumn{5}{l}{\underline{Stellar Mass Tully-Fisher Relation}} \\
$V$ & $M_*$ & $4.70\pm0.22$ & $-0.20\pm0.48$ & 0.21  \\
$I$ & $M_*$ & $5.01\pm0.31$ & $-0.84\pm0.64$ & 0.20   \\
3.6 & $M_*$ & $4.93\pm0.17$ & $-0.61\pm0.36$ & 0.15  \\
\multicolumn{5}{l}{~} \\
\multicolumn{5}{l}{\underline{Baryonic Tully-Fisher Relation}} \\
$V$ & $M_b$ & $3.92\pm0.18$ & $\phm{-}1.81\pm0.39$ & 0.17   \\
$I$ & $M_b$ & $4.02\pm0.24$ & $\phm{-}1.56\pm0.51$ & 0.15  \\
3.6 & $M_b$ & $4.09\pm0.15$ & $\phm{-}1.49\pm0.32$ & 0.13  \\
\multicolumn{5}{l}{~} \\
\multicolumn{5}{l}{\underline{Gas-only Tully-Fisher Relation}} \\
\dots & $M_g$ & $3.24\pm0.28$ & $\phm{-}2.89\pm0.59$ & 0.29 
\enddata
\tablecomments{Fits to the data in Table \ref{basicdata} of the form $\log \ell = \log A + x \log V_f$.}
\label{Fitstats}
\end{deluxetable}

To fit the data here, we use the maximum likelihood method of \citet{benfit}.
This allows for a finite intrinsic scatter in the ordinate in addition to random errors in both dimensions.
The uncertainty that we estimate in the abscissa $V_f$ is given in Table~\ref{basicdata}.
The error in luminosity is dominated by uncertainty in distance rather than in apparent magnitudes.
The uncertainty in the distance to each galaxy is heterogenous and hard to quantify with confidence.
We therefore treat it as a contributor to the scatter.
The best-fit scatter found in this way is a combination of random errors and intrinsic scatter, placing an upper limit on the latter.
The results of the fits are reported in Table~\ref{Fitstats}.

The slope of the LTFR becomes progressively steeper as we go to progressively redder bands (Table~\ref{Fitstats}), 
consistent with previous findings \citep{TFdust,verhTF}.  
As one might expect, the [3.6] LTFR has less scatter and better determined fit coefficients than that in $V$ or $I$.
Nevertheless, we caution against over-interpreting the specifics of these LTFR fits given the high gas content of
the slower rotators. 

The LTFR found here are steeper than many published relations.  
This is expected, as the particular value of the fitted LTFR slope depends on the dynamic range of the data. 
As one goes to lower velocities, one finds more gas rich galaxies.  The luminosity progressively underestimates the mass,
so these galaxies fall too low, steepening the fitted slope.  The scatter in luminosity at a given velocity also goes up \citep{MvDG98}.
The precise value of the fitted slope therefore depends on both the dynamic range of the sample, and the happenstance of the
gas fractions of the galaxies therein.  Magnitude limited samples will be biased to lower slopes, as they will preferentially
include brighter, lower gas fraction galaxies at a given $V_f$.  This can be a strong effect, since it occurs at one end of the relation
where relatively few points can have considerable leverage on the best-fit slope.
In our sample, exclusion of the three slowest rotators (with $V_f \approx 50\;\kms$) flattens the fitted slope perceptibly
(if not significantly). 

An important consequence of sample dependence is that there probably is no such thing as a universally correct LTFR.
The ``right'' answer is unique to each sample.  This effect is difficult to discern in most samples, which tend to lack galaxies
with $V_f \lesssim 90\;\kms$ where the effect becomes pronounced.  This may be enough to explain subtle differences
in published calibrations.  It may also introduce systematic errors in distance determinations, as one is sampling different parts of the relation
at different distances.  Fortunately, this effect will be small as long as only bright galaxies are utilized in both calibration and distance determination.

\subsection{Stellar Mass}  
\label{sec:SMass}

Stellar mass is a more fundamental physical quantity than luminosity.  The LTFR can and does vary from band to band.
These all presumably originate from the same STFR, with differences in the LTFR reflecting systematic variations in the stellar
populations composing fast and slow rotators.  From the color-magnitude relation, one would expect progressively steeper LTFR
as one goes from blue to red bands, but these should all originate from one STFR.

The stellar mass is estimated from $M_* = \ML L$ for each band with the mass-to-light ratio \ML\ estimated by population synthesis models. 
Many models currently in wide use do not pass a straightforward self consistency test, in that stellar masses estimated from the luminosity in
different pass bands for the same galaxy return wildly different results \citep{MS2014}.  By comparing data over a broad range,
\citet{MS2014} constructed empirically self-consistent color-\ML\ relations.

We employ the CMLR of \citet{MS2014} to estimate \ML\ with $B-V$ colors, which 
we found to be the most sensitive to variations in stellar mass-to-light ratio among the available colors.
To be specific, we adopt the model of \citet{Bell03} modified to give self consistent photometric results
as quantified in Table~7 of \citet{MS2014}:
\begin{eqnarray}
\label{eqn:RevBell}
\log \ML^V = -0.628 + 1.305 (B-V) \nonumber \\
\log \ML^I = -0.275 + 0.612 (B-V) \\
\ML^{[3.6]} =  0.47 \nonumber
\end{eqnarray}
As expected, the dependence of \ML\ on color becomes weaker as one goes to redder bands.  It effectively disappears in the NIR,
so we neglect the meaninglessly tiny color slope ($-$0.007) at [3.6].  

We make this particular choice of model because the model of \citet{Bell03} required the least modification of the various models considered.
By construction, the $V$-band \ML-indicator is identical to the original one of \citet{Bell03}, and is subject to the usual normalization uncertainty 
due to the IMF.  Both the $I$-band and [3.6] \ML\ are modified to obtain photometric self-consistency.
Requiring self-consistent stellar masses implies $0.4 < \ML^{[3.6]} < 0.5$ for all models, including those that initially predicted
rather different values.  A constant NIR \ML\ is therefore a good approximation.  

The dependence of the NIR \ML\ on color in eq.~\ref{eqn:RevBell} is considerably weaker than the already weak relation anticipated in many models.  
This is required for self-consistency:  the photometric data are not consistent with NIR mass-to-light ratios that depend much on color
\citep[see][]{MS2014}.  While there can certainly be some color dependence, the variation of $\ML^{[3.6]}$ with color appears to
be rather less than the intrinsic scatter in $\ML^{[3.6]}$ from galaxy to galaxy, which itself is also fairly small. 

In practice, the weak relation between NIR \ML\ and color means that NIR luminosity is a good proxy for stellar mass.
One might hope to do better by fitting the entire SED of a galaxy, and this no doubt matters for extreme objects like those
dominated by the light of very young stellar populations.  However, for most galaxies, there is no value added from this
procedure.  Most of the information about \ML\ is contained in the first color ($B-V$ here), which is already negligible in the NIR. 
Further corrections from other colors carry progressively less information.  
In the case of [3.6] and the $K_s$-band, attempts at SED fitting are more likely to add noise than improve the estimate of \ML.  
Worse, they are prone to induce systematic errors from imperfections in the population models.  
For example, many models assume a single metallicity for the stellar population.  This appears to
be one driver for the predicted dependence of the NIR mass-to-light ratio on color being stronger than observed \citep{MS2014}.
Models with realistic metallicity distributions have a much weaker dependence of NIR \ML\ on color \citep{SM14pop}.

\begin{figure*}
\epsscale{1.0}
\plotone{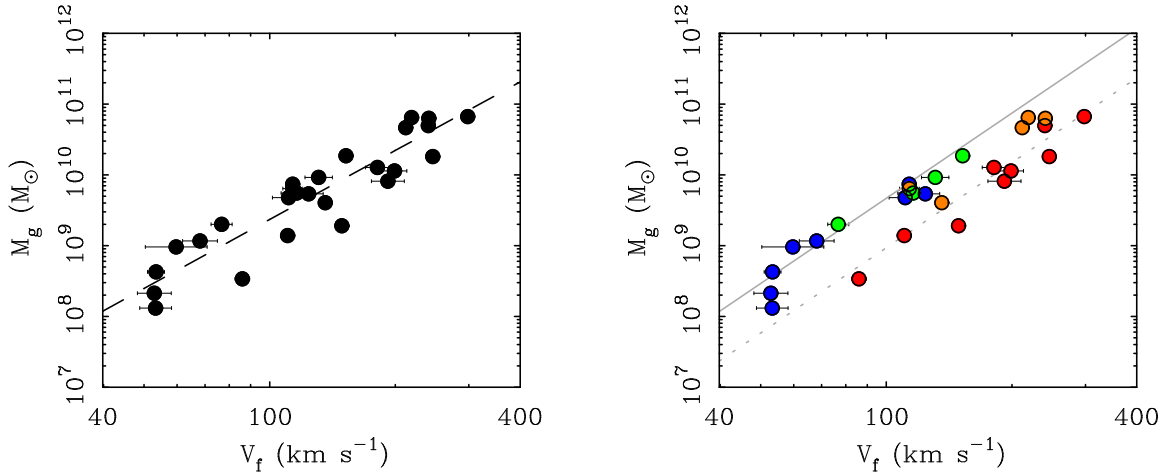}
\caption{The gas-only Tully-Fisher relation.  Gas mass is plotted against rotation velocity without any reference to stellar luminosity.
The data are plotted and fit (dashed line) in the left panel (see Table~\ref{Fitstats} for coefficients).
The fit looks reasonable, though the scatter is larger than in any of the LTFR.  
The right panel shows the same data, but with points color coded by gas fraction.
To estimate the ratio $M_*/M_g$, we use $M_* = 0.47 L_{[3.6]}$.
Red points have $M_*/M_g > 2$, orange points 
$1 < M_*/M_g < 2$, green points $\onehalf < M_*/M_g < 1$, and blue points $M_*/M_g < \onehalf$.
The data clearly segregate by gas fraction.  Gas rich galaxies follow the BTFR (solid grey line)
while gas poor galaxies fall well below it.  For comparison, the dotted lines parallels the BTFR but is shifted by a factor of 
five.  This is the amplitude of the shift one would expect for a typical Milky Way-like spiral with a $\sim 20\%$ gas fraction.
The segregation by gas fraction renders meaningless the fit in the left panel.
\label{gasTF}}
\end{figure*}

If the stellar masses-to-light ratios estimated with eq.~\ref{eqn:RevBell} are correct, the \ML\ for each band should succeed in converting
the different LTFR observed for each independent band into the same STFR.  This procedure is indeed successful.  
The STFR in Table~\ref{Fitstats} are equivalent within the errors.

Since the adopted NIR mass-to-light ratio is constant, the [3.6] STFR is identical to the [3.6] LTFR except for the change in normalization
due to the non-unity value of the mass-to-light ratio.  Consequently, it is the slope of the $V$ and $I$ LTFR that change to result
in a STFR that matches that from [3.6].  This goes in the expected sense from the color-magnitude relation: fainter galaxies are
bluer and have lower \ML\ on average.  Use of incorrect CMLR would lead to discordant STFR in the various bands.
That we find consistent SFTR provides independent confirmation that the CMLR in 
eq.~\ref{eqn:RevBell} determined photometrically by \citet{MS2014} are very nearly correct.

The conversion from the LTFR to the STFR can only work to the extent that $B-V$ is predictive of \ML.  
That it does work suggests that $B-V$ is an adequate \ML\ indicator.
It certainly is not perfect: one expects a fair amount of scatter about the mean CMLR.
The expected consequence is an increase in the scatter in the STFR relative to the LTFR.  Even though the mean slope
of the CMLR succeeds in converting the LTFR into the STFR, the scatter in the latter increases due to misestimates
of the mass-to-light ratios of individual galaxies.  That is, the scatter in the STFR should include whatever scatter is 
already in the LTFR, plus an extra component for the scatter in the CMLR.  This effect does appear to be present in the data, 
as seen by the amount of scatter in Table~\ref{Fitstats}.  

From a population perspective, one expects a greater amount of scatter in \ML\ at bluer wavelengths.
The scatter fit in Table~\ref{Fitstats} is a combination of intrinsic scatter and that from random errors like those in distance.
While there is considerable uncertainty in these uncertainties, we estimate that random errors contribute $\sim 0.08$ dex
to the scatter in the ordinate of the various Tully-Fisher relations.  Subtracting this in quadrature from the fitted value for each
band gives an estimate of the intrinsic scatter in \ML: 0.16 dex in $V$, 0.13 in $I$, and $0.12$ in [0.36].
For comparison, \citet{BdJ01} anticipate 0.1 dex of scatter in $K$ from variations in the star formation history alone,
and \citet{Port04} anticipate 0.15 dex. 

It thus appears that the observed scatter is entirely explained by random errors plus the expected scatter in mass-to-light ratios.
The scatter in \ML\ is consistent with that expected from the inevitable variations in star formation histories.  This leaves essentially no room
for variations in the IMF from galaxy to galaxy.  Apparently, the IMF in disk galaxies is universal, at least averaged over the many
star formation events that build a galactic disk over a Hubble time.

There is also precious little room for intrinsic scatter in whatever physics underlies the Tully-Fisher relation.
Given the small size of the sample, we should not over-interpret the precise values of the scatter.   However, it is not obvious
that larger samples will inevitably lead to larger scatter, as they bring with them the larger risk of unaccounted systematic errors
\citep[see discussion in][]{M12}.  Bear in mind that in order to increase the intrinsic scatter of this sample, we must have
\textit{overestimated} random errors.  Even in the limit of zero experimental error and no scatter in \ML, 
the intrinsic scatter is still small ($< 0.15$ dex; Table~\ref{Fitstats}).

\subsection{Gas Mass}  
\label{sec:GMass}

The atomic and molecular gas masses given in Table~\ref{basicdata} are used to compute the total gas mass
\begin{equation}
M_g = 1.33[M(HI)+M(H_2)].
\label{eqn:gasmass}
\end{equation}
The factor 1.33 accounts for the mass in helium.  
One can debate the appropriate value of the second digit past the decimal, but not at a level that matters here.
We adopt 1.33 for consistency with \citet{M12}, whose BTFR calibration will be utilized in \S \ref{sec:btfrmass}.
The contribution of other gas phases is generally negligible, at least within the visible disk where all relevant quantities are measured.

Luminosity is the original ordinate of the Tully-Fisher relation. 
Stars are the dominant baryonic mass component of bright galaxies.
We therefore do not expect there to be a meaningful gas-only Tully-Fisher relation.
For completeness, we do construct one.  It is shown in Fig.~\ref{gasTF}.

At first glance, the relation in Fig.~\ref{gasTF} appears perfectly reasonable.
It has considerably greater scatter than any of the STFR, but it is certainly possible to fit a line through the data (left panel of Fig.~\ref{gasTF}).
This is not particularly meaningful, as the scatter here is real.  Galaxies of the same $V_f$ can differ in gas mass by an order of magnitude.

Galaxies segregate by gas fraction (right panel of Fig.~\ref{gasTF}).
To estimate the gas fraction, we use $\ML^{[3.6]} = 0.47$ from eq.~\ref{eqn:RevBell} to estimate the stellar mass, 
and bin the data by the ratio $M_*/M_g$.  Galaxies with $M_* < M_g$ hover near the BTFR previously fit to gas rich galaxies \citep{M12}.
Since stellar mass is ignored here, these should all fall slightly below the BTFR line.  Most do, though a few fall a tad too high,
albeit within the errors.  In contrast, galaxies with $M_* > M_g$ all fall well below the line.  The obvious inference is that we are
missing an important component of the mass in these galaxies:  the stars.  Indeed, the amount of the shift is what we expect
for the typical gas fractions of bright spirals.  

Just as we should not ignore the gas in faint galaxies, we should not ignore the stars in bright ones.
Neither the LTFR nor the gas-only Tully-Fisher relation are fundamental.
Instead, they must stem from a more basic relation that does not distinguish between mass in the form of stars or gas.

\subsection{Baryonic Mass}
\label{sec:BMass}

In order to construct the BTFR, we take the sum of stellar and gas mass to obtain the baryonic mass:
\begin{equation}
M_b = M_* + M_g.
\label{eqn:barymass}
\end{equation}
The stellar mass is computed with eq.~\ref{eqn:RevBell} and the gas mass is from eq.~\ref{eqn:gasmass}.
The BTFR resulting from the stellar mass as computed from each of $V$, $I$, and [3.6] is shown in the bottom row of Fig.~\ref{TF}.

If all has gone well, this BTFR should be consistent from band to band.
This should necessarily follow if the STFR has been computed correctly for each band.
Indeed, this is the case:  the relations computed separately for each band are mutually consistent.
Both the slopes and the intercepts are indistinguishable within the errors.
This convergence is readily seen in Fig.~\ref{TF}.
These relations are also consistent with the BTFR previously fit to independent data for gas rich, predominantly slow rotators by \citet{M12}.

Another encouraging point is that scatter in the BTFR is less than that in either the LTFR or the STFR.
Including the gas mass as well as the stellar mass has made a better relation in this sense.
Indeed, the relation is near perfect in the sense that their appears to be very little intrinsic scatter.
In \S \ref{sec:SMass} we estimated the intrinsic scatter in stellar mass-to-light ratios for this sample to be
0.16 dex in $V$, 0.13 in $I$, and $0.12$ in [0.36].  The scatter in the BTFR is 
0.17 dex in $V$, 0.15 in $I$, and $0.13$ in [0.36] (Table~\ref{Fitstats}).  
These are effectively identical given the uncertainties, leaving
precious little room for intrinsic scatter in the relation itself.

The fundamental, underlying physical relation appears to be one between baryonic mass and asymptotic rotation velocity.
The data are consistent with a single power law (a single linear slope in log-log space) over four decades in baryonic mass.
There is no evidence for a bend or break in the relation as sometimes anticipated by models \citep[e.g.,][]{TGKPR}.

The luminosity of the traditional LTFR is merely a proxy for baryonic mass.  This works well for bright spirals that are gas poor,
but starts to break down at lower velocity where galaxies tend to be more gas rich.  As a consequence, the LTFR suffers increased
scatter at low rotation velocity, and may appear to bend.  This simply illustrates the shortcomings of luminosity as a mass indicator,
not any break in the underlying relation.  

Since luminosity is an imperfect indicator of baryonic mass, calibrations of the LTFR are inevitably inaccurate at some level.
Bright spirals do contain gas, and the gas fraction varies from galaxy to galaxy.  The LTFR will inevitably deviate from the underlying BTFR
in a way that depends both on the band pass and the gas fractions of the sample of calibration galaxies.  Fortunately this appears to
be a minor effect for bright galaxies.

The small intrinsic scatter in the BTFR suggests that there is no large reservoir of missing baryons remaining to be discovered in
the disks of spiral galaxies \citep{btforig}.  It is obvious if we miss either the stars or the gas (Figures \ref{TF} and \ref{gasTF}).
If we were still missing a comparable number of baryons, then the sum in eq.~\ref{eqn:barymass} would be incomplete.  
This would show up as a systematic deviation from the BTFR, which is not apparent.  One can always have
\textit{some} mass in an undetected component \citep[e.g., in the form of very cold molecular gas --- ][]{pfennBTF}, 
but its mass must be $\ll M_b$ from eq.~\ref{eqn:barymass} so that the data stay
within the scatter.  Consequently, the stars and cold gas that we detect appear to compose the bulk of the baryonic mass in disk galaxies.

\section{Stellar Masses from the BTFR}
\label{sec:btfrmass}

Here we employ the BTFR as a method to estimate stellar mass.
Rather than trust population synthesis models, we ask what the stellar mass must be in order for a galaxy to lie on the BTFR.
This would obviously be a completely circular exercise if we were to use the BTFR calibrated with star dominated galaxies as in the previous section.
Fortunately, recent work on gas dominated galaxies \citep{begum,swaterswhisp,stark,trach} has made it possible to obtain an absolute calibration of the mass scale of the 
BTFR that is effectively independent of stellar mass estimates \citep{M12}.  

\begin{deluxetable*}{lccccc}
\tablewidth{0pt}
\tablecaption{Stellar Masses and Mass-to-Light Ratios from the BTFR}
\tablehead{
\colhead{Galaxy} & \colhead{$M_*$[3.6]\tablenotemark{a}} 
& \colhead{$M_*$(BTFR)} & \colhead{$\ML^V$} & \colhead{$\ML^I$} & \colhead{$\ML^{[3.6]}$} \\
& \multicolumn{2}{c}{($10^9\;\mathrm{M}_{\sun}$)} & \multicolumn{3}{c}{(\MLsun)} 
 }
\startdata
DDO 154	 &0.0341  &$ 0.248 \pm0.142$	&$ 4.81\pm2.76$	&$ 8.44\pm4.84$	&$ 3.42\pm1.96$ \\
D631-7	 &0.0458  &$ 0.156 \pm0.147$	&$ 2.89\pm2.72$	&$ 3.48\pm3.27$	&$ 1.60\pm1.51$ \\
DDO 168	 &0.0881  &$-0.0431\pm0.0866$	&$-0.26\pm0.53$	&  \dots       	&$-0.23\pm0.46$ \\
D500-2	 & 0.186  &$-0.158 \pm0.413$	&$-0.52\pm1.35$	&$-0.70\pm1.83$	&$-0.40\pm1.05$ \\
NGC 2366	 & 0.336  &$-0.365 \pm0.419$	&$-0.80\pm0.92$	&$-0.99\pm1.13$	&$-0.51\pm0.58$ \\
IC 2574	 &  1.10  &$-0.373 \pm0.435$	&$-0.36\pm0.42$	&$-0.39\pm0.46$	&$-0.16\pm0.19$ \\
F563-1	 &  1.34  &$ 2.36  \pm2.73$	&$ 2.09\pm2.42$	&$ 1.89\pm2.19$	&$ 0.83\pm0.95$ \\
NGC 2976	 &  1.50  &$ 2.23  \pm0.53$	&$ 1.98\pm0.47$	&$ 2.13\pm0.51$	&$ 0.70\pm0.16$ \\
F568V-1	 &  1.97  &$ 5.71  \pm3.85$	&$ 2.42\pm1.64$	&$ 2.54\pm1.71$	&$ 1.36\pm0.92$ \\
NGC 1003	 &  2.78  &$ 0.41 \pm1.12$	&$ 0.15\pm0.40$	&  \dots       	&$ 0.07\pm0.20$ \\
F568-1	 &  3.33  &$ 2.98  \pm3.13$	&$ 0.96\pm1.01$	&$ 1.00\pm1.06$	&$ 0.42\pm0.44$ \\
NGC 7793	 &  3.55  &$ 5.59  \pm1.43$	&$ 1.87\pm0.48$	&$ 3.10\pm0.79$	&$ 0.74\pm0.19$ \\
UGC 128	 &  5.30  &$ 4.62  \pm4.58$	&$ 1.03\pm1.02$	&$ 1.10\pm1.09$	&$ 0.41\pm0.41$ \\
NGC 2403	 &  5.71  &$ 11.9  \pm2.8$	&$ 3.08\pm0.72$	&$ 2.38\pm0.55$	&$ 0.98\pm0.23$ \\
NGC 925	 &  7.42  &$ 1.42  \pm2.00$	&$ 0.17\pm0.24$	&$ 0.17\pm0.24$	&$ 0.09\pm0.12$ \\
NGC 3198	 &  14.7  &$ 21.2  \pm3.9$	&$ 1.71\pm0.32$	&$ 1.34\pm0.25$	&$ 0.68\pm0.13$ \\
NGC 3621	 &  15.5  &$ 6.61  \pm3.86$	&$ 0.98\pm0.57$	&$ 0.93\pm0.54$	&$ 0.20\pm0.12$ \\
NGC 3521	 &  42.8  &$ 55.6  \pm24.8$	&$ 3.57\pm1.59$	&$ 2.40\pm1.07$	&$ 0.61\pm0.22$ \\
NGC 3031	 &  49.6  &$ 62.3  \pm22.5$	&$ 2.70\pm0.97$	&$ 1.61\pm0.58$	&$ 0.59\pm0.26$ \\
NGC 5055	 &  63.5  &$ 37.8  \pm14.8$	&$ 1.44\pm0.56$	&$ 0.94\pm0.37$	&$ 0.28\pm0.11$ \\
NGC 2998	 &  75.0  &$ 47.9  \pm13.0$	&$ 0.64\pm0.17$	&  \dots       	&$ 0.30\pm0.08$ \\
NGC 6674	 &  108.  &$ 94.1  \pm22.9$	&$ 1.49\pm0.36$	&  \dots       	&$ 0.41\pm0.10$ \\
NGC 7331	 &  122.  &$ 153.  \pm32.$	&$ 4.00\pm0.84$	&$ 2.28\pm0.47$	&$ 0.59\pm0.12$ \\
NGC 801	 &  126.  &$ 42.9  \pm15.4$	&$ 0.60\pm0.21$	&  \dots       	&$ 0.16\pm0.06$ \\
NGC 5533	 &  142.  &$ 106.  \pm24.$	&$ 1.69\pm0.38$	&  \dots       	&$ 0.35\pm0.08$ \\
UGC 2885	 &  220.  &$ 304.  \pm69.$	&$ 1.70\pm0.38$	&  \dots       	&$ 0.65\pm0.15$ 
\enddata
\tablenotetext{a}{Stellar masses assuming $\ML^{[3.6]} = 0.47\;\MLsun$ are included for comparison.}
\label{BTFmasses}
\end{deluxetable*}

The ideal situation would be a calibration of the BTFR with galaxies made purely of gas.  In practice, this is impossible. 
Nevertheless, it is possible to select galaxies that are sufficiently gas rich that the 
error in the stellar mass becomes a minor contributor to the error budget.  This occurs when $M_g > M_*$ \citep{M11}.  

\citet{M12} discusses fits to the BTFR for several distinct samples of gas rich galaxies.
Combined, these are consistent with the BTFR  
\begin{equation}
\label{eqn:gasrichBTFR}
\log M_b = 1.67 + 4 \log V_f.
\end{equation}
In order to estimate the stellar mass, we assume that star dominated galaxies exactly follow eq.~\ref{eqn:gasrichBTFR}.
This seems like a reasonable assumption given the results of the previous section.
However, the gas rich galaxies fit by \citet{M12} are much slower rotators than the sample of objects here (Fig.~\ref{6decades}).
It is therefore a big extrapolation to apply eq.~\ref{eqn:gasrichBTFR} to the bright galaxies in Table~\ref{basicdata}.

\begin{figure*}
\epsscale{1.0}
\plotone{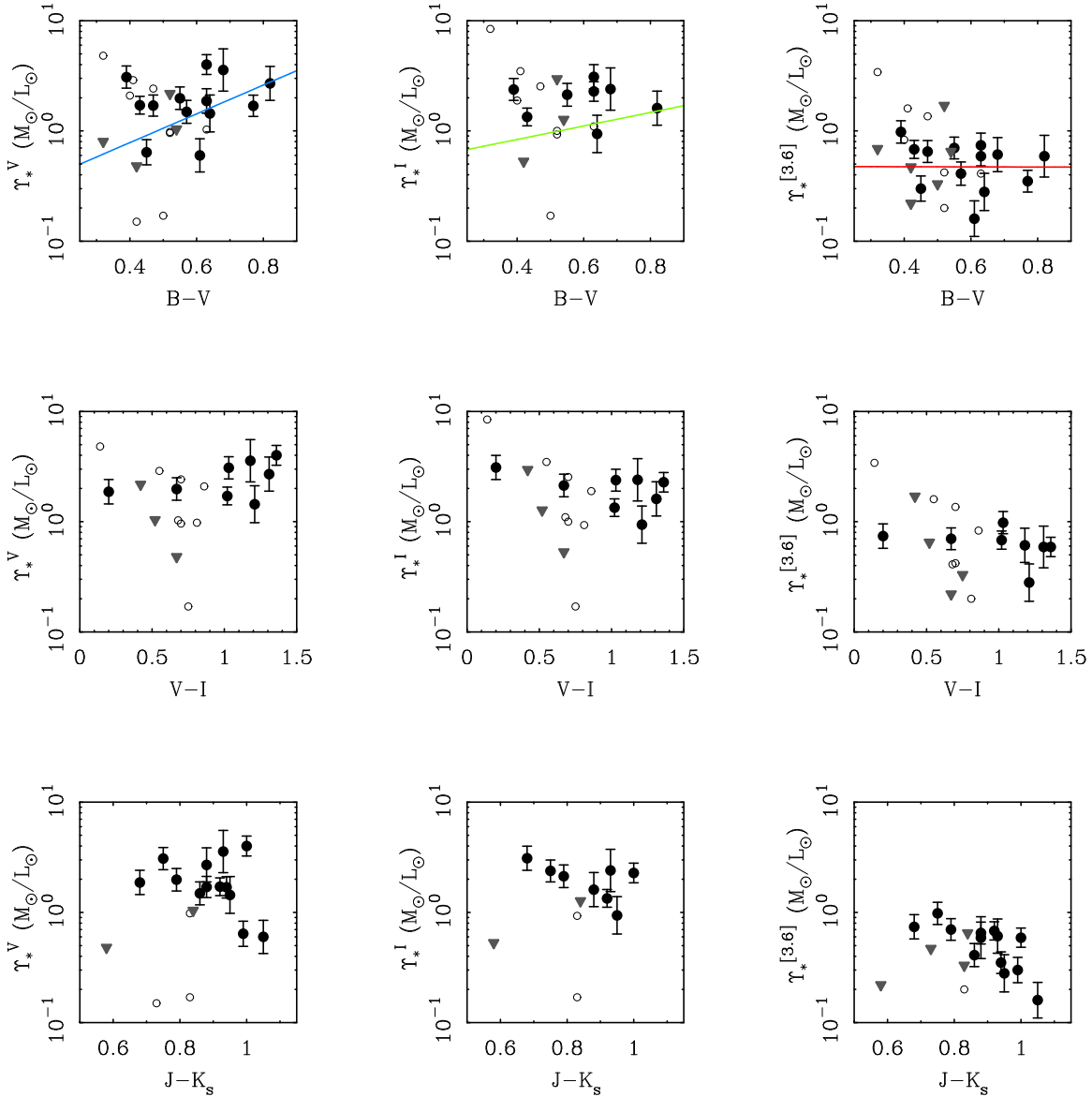}
\caption{Stellar mass-to-light ratios (Table~\ref{BTFmasses}) inferred from the gas rich calibration of the BTFR. 
Mass-to-light ratios in $V$ (left column), $I$ (middle column), and [3.6] (right column) are plotted 
as a function of color in $B-V$ (top row), $V-I$ (middle row), and $J-K_s$ (bottom row).
Data with uncertainties in stellar mass $< 50\%$ are shown as solid points.  Less accurate data are shown as open circles.
Galaxies with formally negative stellar masses are shown as $2 \sigma$ upper limits (grey triangles).
The mass-to-light ratios predicted by eq.~\ref{eqn:RevBell} are shown as lines in the panels of the top row. 
\label{MLBV}}
\end{figure*}

We first use the measured rotation velocity and eq.~\ref{eqn:gasrichBTFR} to estimate of the total baryonic mass.
The stellar mass then follows by subtracting the gas mass:
\begin{equation}
M_* = M_b - M_g.
\label{eqn:Mst}
\end{equation}
The results of this computation are reported in Table~\ref{BTFmasses}.
The first column of Table \ref{BTFmasses} gives the name of each galaxy, as in Table~\ref{basicdata}.
Column 2 gives the stellar mass as estimated by the [3.6] luminosity assuming $\ML^{[3.6]} = 0.47\;\MLsun$.
This is provided for direct comparison to the stellar mass estimated with eq.~\ref{eqn:Mst}, which is given in
column 3. The formal uncertainty in the stellar mass is estimated by propagating the uncertainty in $V_f$, distance, and the gas rich 
BTFR calibration.  The corresponding mass-to-light ratios in the $V$, $I$, and [3.6] bands are given in columns 4--6.

\begin{figure*}
\epsscale{1.0}
\plotone{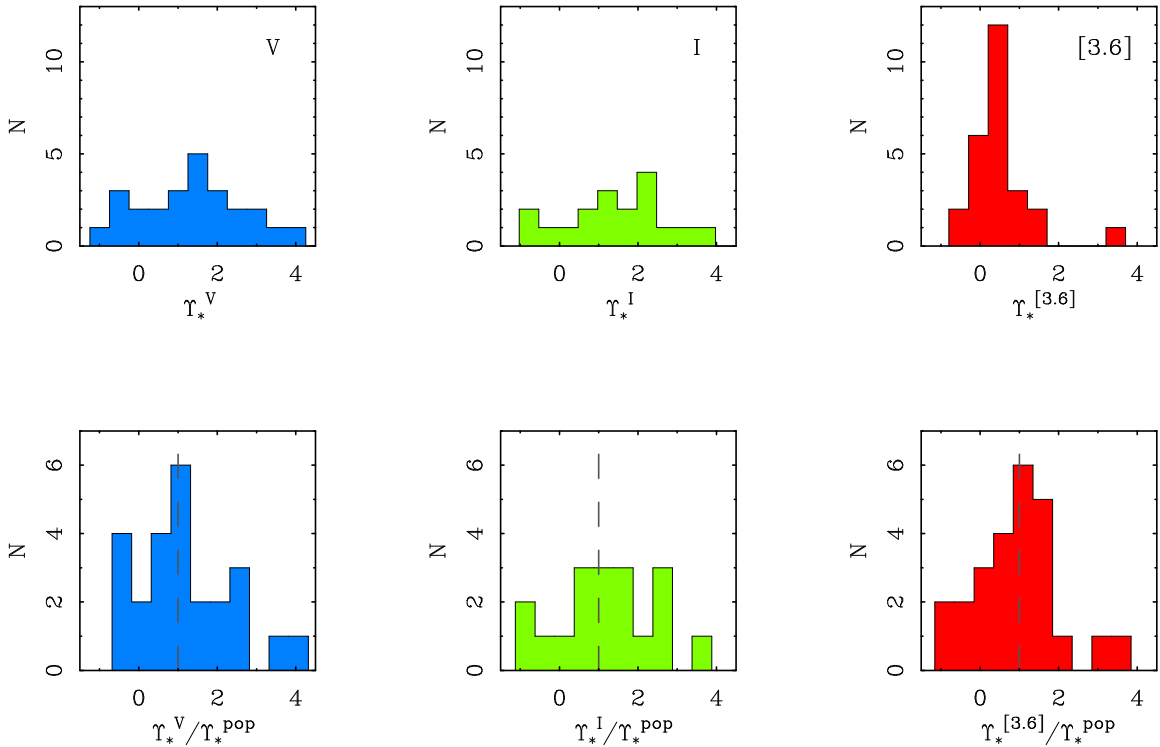}
\caption{Histograms of stellar mass-to-light ratios inferred from the BTFR in $V$ (left column), $I$ (middle column), and [3.6] (right column).
The top row shows the measured mass-to-light ratios while the bottom row shows \ML\ relative to the population synthesis 
expectation of eq.~\ref{eqn:RevBell} (dashed lines).  The bin size and plot scale are the same for all filters within each row.
The data are consistent with $\ML^{[3.6]} = 0.45\;\MLsun$ with modest intrinsic scatter.
\label{MLhist}}
\end{figure*}

Estimating stellar masses and mass-to-light ratios with this procedure is very demanding on data quality.
Starting from $V_f$ which is typically of order $\sim 100\;\kms$, we first raise this number to a high power.  We then multiply by
an empirically determined constant of the same order.  This gives a large number appropriate to the baryonic mass of a galaxy.  
We then subtract another large number, the gas mass.  This procedure invites the propagation of errors.  
The stellar masses we obtain are nevertheless quite reasonable (Table~\ref{BTFmasses}).
We should, however, not be surprised if this method sometimes returns unphysical results, like negative stellar masses. 

Indeed, negative stellar masses are inferred four times, in DDO 168, D500-2, NGC 2366, and IC 2574.
These four galaxies are the points that lie slightly above the line in Fig.~\ref{gasTF}.  
In no case is the stellar mass significantly negative: all four are within $1 \sigma$ of being positive.

The opposite situation also happens.  
DDO 154 is inferred to have an unreasonably large mass-to-light ratio --- it is the lowest point in Fig.~\ref{gasTF}.  
However, its stellar mass is non-zero by less than $2 \sigma$, so its large inferred mass-to-light ratio is not meaningful.  

The few extremal cases are an indication of the success of the procedure.
If we attempt the same exercise using line-widths instead of resolved rotation curves, we routinely find negative masses.
We often also find super-maximal mass-to-light ratios, which do not occur in Table~\ref{BTFmasses}.
The lower scatter in the BTFR obtained with resolved rotation curves is essential to the method.

One should nevertheless treat the results for individual galaxies with caution.  The data are statistically indicative, but it is very hard to have confidence in
any one stellar mass.  Given the large extrapolation from the gas rich regime, the stellar mass of the brightest galaxies 
is extremely sensitive to the slope of the BTFR (see \S \ref{sec:dmbtfr}).  

We can check the inferred mass-to-light ratios against the expectations of population synthesis models.
This is done in Figure \ref{MLBV}, where we show the mass-to-light ratios from Table~\ref{BTFmasses} as a function of color.  
Figure \ref{MLhist} shows histograms of the BTFR mass-to-light ratios.

The mass-to-light ratios estimated from the BTFR are consistent in amplitude with that expected from population synthesis.
This was not guaranteed:  the BTFR calibrated by gas rich galaxies provides a constraint that is independent from the population models.
While it was clear already from Fig.~\ref{TF} that this would follow, it need not have.  Indeed, it would not have if we employed a model
much different from that of eq.~\ref{eqn:RevBell}, of which there are many examples in the literature.  

Nevertheless, the amplitude, slope, and scatter in \ML\ expected form population synthesis are consistent with the gas rich BTFR.
While the data are consistent with the CMLR determined photometrically, they are not precise enough to further constrain it.
That is, we cannot use these data to measure the color slopes of the CMLR.

Indeed, Fig.~\ref{MLBV} does not show a strong dependence of \ML\ on color.  
As expected, the dependence of \ML\ on $B-V$ becomes weaker as one goes from optical to NIR.  
There is no hint of a slope at all at [3.6].  As expected from a population perspective, the NIR \ML\ from the BTFR is effectively constant.

If the dependence of \ML\ on $B-V$ is weak, it is virtually non-existent for the other colors.
There is a hint in the data that $\ML^V$ might become larger for very red $V-I$.
Otherwise, \ML\ does not appear to be particularly sensitive to either $V-I$ or $J-K$.
This is not surprising.  It was already clear from the photometry \citep{MS2014} that $B-V$ contained most of the information about \ML,
with $V-I$ carrying only a small additional amount beyond that.  There is no useful information about \ML\ in $J-K$ \citep[see also][]{SM14pop}.

The histogram of \ML\ (Fig.~\ref{MLhist}) shows scatter about a typical value.
The widths of the histograms basically reflect the propagation of errors: the data are consistent with a narrow intrinsic scatter (\S \ref{sec:BMass}).
Nevertheless, it is worth noting that there is a clear and well defined peak at a preferred value in the NIR: $\ML = 0.45\;\MLsun$.
 
The top row of Fig.~\ref{MLhist} shows the raw distribution of \ML, and does not consider the expected variation with color.  
Since there is no expected variation with color in [3.6], the peak is clearer there than it can be in either $V$ or $I$.
In the bottom row of Fig.~\ref{MLhist}, we show the histogram of \ML\ relative to that expected from eq.~\ref{eqn:RevBell}.  
Here it can be seen that the data do peak near the value expected from population synthesis:  there is value in the 
color informed \ML\ estimate of the CMLR.

An obvious consequence is that the luminosity in any band is a good approximation of the stellar mass.
This approximation can be improved with use of color information.  The amount of improvement is limited by the intrinsic scatter in the CMLR.
Both the scatter and the color term are minimized in the NIR.  This makes the NIR luminosity the best available proxy for stellar mass.

The results here are consistent with the results of \S \ref{sec:BMass}.  The stellar population estimate of the BTFR gives a relation indistinguishable
from the separate gas rich calibration. It makes little difference which method we choose.  Either the data fall exactly on the BTFR and give
the scatter seen in Fig.~\ref{MLBV}, or the galaxies follow exactly this lines of eq.~\ref{eqn:RevBell} in Fig.~\ref{MLBV} with the scatter
seen in Fig.~\ref{TF}.  There has to be some scatter in \ML, leaving precious little room for intrinsic scatter in the BTFR.

It remains unclear why the BTFR is so well posed as a relation between baryonic mass and the flat rotation velocity.
It does not follow from structure formation simulations without further assumptions.  
It has become common to invoke feedback from star formation in this context \citep[e.g.,][]{fabioTF,Dutton2012}, 
but it is hard to see how the chaotic processes of supernovae returning energy to the ISM can result in the negligible scatter observed in the BTFR.  
The only theory to anticipate the specific $M_b$-$V_f$ relation that we observe is MOND \citep{milgrom83}.

Regardless of the physics underlying the BTFR, it is of enormous utility to have such a relation.
We emphasize that the BTFR constructed here is empirical, making no reference to either dark matter or MOND.
It relies only on observational data and stellar population synthesis models to inform the CMLR.

\section{A Constraint on the IMF}
\label{sec:IMF}

Returning to the matter of stellar mass, there is additional information in the BTFR.  
The normalization of the BTFR calibrated by gas rich galaxies is effectively independent of stellar mass estimates.  
In contrast, the population synthesis mass estimates are very sensitive to the IMF.  That the two are consistent
implies that we are not far wrong about the IMF.  Nevertheless, we can query the data to see what mass-to-light ratio is preferred.
Rather than take the population synthesis mass at face value and fit the BTFR, we adjust the stellar mass of bright galaxies
to find the best agreement with the BTFR previously fit to gas rich galaxies.
This leads to a constraint on the IMF \citep{stark}.

In order to find the ideal match with the BTFR, we adjust the zero point in eq.~\ref{eqn:RevBell} while holding the color slope fixed.
We then find the zero point that best matches the BTFR of \citet{M12}.   That is, we determine the value $\Delta a$ in
\begin{equation}
\label{eqn:adjustML}
\log \ML = a + b (B-V)  + \Delta a
\end{equation}
that best matches the gas rich BTFR.

The net effect of this exercise is a small shift of the intercept in each band.  
The result is $\Delta a = -0.015$ in $V$, $+0.06$ in $I$, and $-0.019$ in [3.6].
This is what it takes to make the dashed lines in the bottom row of Fig.~\ref{TF} fall on top of the solid lines.

These shifted zero point are not significantly different from the original population synthesis values.
For example, the NIR value changes from $\ML^{[3.6]} = 0.47$ to $0.45\;\MLsun$.
Similarly, in the galaxies for which we have $K$-band data, the mean shifts from $\ML^{K} = 0.60$ to $0.57\;\MLsun$
($\Delta a = -0.022$).  This corresponds to a ratio $\ML^K/\ML^{[3.6]} = 1.27$, consistent with the 1.29 we estimated photometrically \citep{MS2014}.

This nice consistency may seem mundane, but is actually a rather important observation.  
That $\Delta a$ is small implies that the population models are nearly correct, provided they adopt the right IMF.
The data are consistent with a Kroupa or Chabrier IMF.  The Salpeter IMF, as usually employed, is a bit too 
heavy \citep[see discussion in][]{stark}.  This is consistent with what we know about the IMF locally, in which the numbers of low mass
stars does not continue to increase as a Salpeter power law \citep{kroupa,chabrier}.

\section{Comparison to Other Methods}
\label{sec:othermeth}

The absolute calibration of stellar mass is of obvious interest.
In this section, we compare the normalization found from the BTFR to other methods in the literature.
A largely consistent picture emerges, with one important exception.

Table~\ref{MLresults} summarizes various estimates of the stellar mass-to-light ratio of disk galaxy stellar populations.
Column 1 gives a name for each method considered as described in column 2.  
Columns 3--6 give typical mass-to-light ratios in the $V$, $I$, $K$, and [3.6] bands.  
We include the $K$-band here for comparison with [3.6].  These two bands are nearly interchangeable photometrically \citep{SM2014,SM14pop}.
We convert between them assuming the ratio of mass-to-light ratios is constant so that results limited to one or the other band can be directly compared.
Column 8 gives the reference for each \ML\ calibration.

\begin{deluxetable}{llccccr}
\tablewidth{0pt}
\tablecaption{Mass-to-Light Ratios Estimates}
\tablehead{
\colhead{Calibration} & \colhead{Method} & \colhead{$\langle \ML^V \rangle$} 
& \colhead{$\langle \ML^I \rangle$}  & \colhead{$\langle \ML^K \rangle$}  & \colhead{$\langle \ML^{[3.6]} \rangle$} & \colhead{Ref.}
}
\startdata
Gas Rich    & BTFR & 1.38 & 1.42 & 0.57 & 0.45 & 1 \\
DiskMass\tablenotemark{a} & Vertical $\sigma_z$  & 0.74 & 0.65 & 0.31 & 0.24 & 2 \\
DiskMass\tablenotemark{a} & DM-adjusted  &0.57 & 0.50 & 0.24 & 0.19 & 3 \\
Milky Way & Star counts  & 1.5\phn & 1.2\phn & \dots & \dots & 4 \\
Milky Way & Terminal $v_t$  &1.53 & 1.22 & 0.63 & 0.49 & 5 \\
Milky Way & Vertical $K_z$  &1.64 & 1.31 & 0.66 & 0.51 & 6 \\
LMC     & Star counts &   \dots & \dots & 0.65 & 0.5\phn & 7 \\
Bell & Popsynth &  1.43 & 1.25 & 0.73 & 0.62 & 8 \\
Portinari &  Popsynth &  1.32 & 1.11 & 0.50 & 0.41 & 9 \\
Zibetti &  Popsynth\tablenotemark{b} &  1.07 & 0.76 & 0.21 & 0.14 & 10 \\
Into & Popsynth\tablenotemark{b} &  1.19 & 0.99 & 0.41 & 0.33 & 11 \\
Eq.~\ref{eqn:RevBell} & Popsynth &  1.43 & 1.24 & 0.60 & 0.47 & 12 \\
Schombert & Popsynth &  \dots & \dots & 0.65 & 0.5\phn & 13 \\
Meidt     & Popsynth &  \dots & \dots & 0.77 & 0.6\phn & 14 
\enddata
\tablerefs{1.~This work. 2.~\citet{diskmass7}. 3.~\citet{diskmassSB}. 
5.~\citet{M08}. 4.~\citet{flynn}. 6.~\citet{bovyrix}. 7.~\citet{Eskew}. 
8.~\citet{Bell03}. 9.~\citet{Port04}. 10.~\citet{Zib09}. 11.~\citet{IP13}.
12.~\citet{MS2014}. 13.~\citet{SM14pop}. 14.~\citet{Meidt2014}. 
}
\tablecomments{Color dependent population synthesis models are evaluated at $B-V = 0.6$.
To compare the $K_s$ and [3.6] bands we assume $\ML^K = 1.29 \ML^{[3.6]}$ \citep{MS2014}, 
with the exception of the BTFR method which independently suggests $\ML^K = 1.27 \ML^{[3.6]}$ (see text).
}
\tablenotetext{a}{For comparison with other methods, the $V$ and $I$-band values are approximated by scaling from 
the $K$-band measurement with eq.~\ref{eqn:RevBell}.}
\tablenotetext{b}{These models do not give self-consistent stellar masses unless the NIR mass-to-light
ratio is adjusted to larger values: $\ML^K \approx 0.58$ and $\ML^{[3.6]} \approx 0.45\;\MLsun$ \citep[see discussion in][]{MS2014}.}
\label{MLresults}
\end{deluxetable}

Our result from \S \ref{sec:btfrmass} is given first.  These are consistent with our previous attempt to employ this method \citep{stark}.
It is also consistent with many, but not all, population synthesis estimates. The NIR \ML\ are hardly different from eq.~\ref{eqn:RevBell}.
The tweak in the $V$-band is also negligible, but that in the $I$-band is more noticeable.  These are sensitive to color, and
are estimated at $B-V = 0.6$ to provide a uniform comparison.  We do not consider the difference between the BTFR $\ML^I \approx 1.4\;\MLsun$
and the value from eq.~\ref{eqn:RevBell} of $\ML^I \approx 1.2\;\MLsun$ to be particularly significant.
It perhaps serves more as an illustration of just how sensitive the method is, as this is the shift ($\Delta a = 0.06$)
required to match the two nearly indistinguishable lines in the bottom middle panel of Fig.~\ref{TF}.
Indeed, the positive $\Delta a$ in the $I$-band (when the other filters are slightly negative) 
may simply be the result of missing data in that filter for some of the brighter galaxies. \\

\subsection{The DiskMass Survey}
\label{sec:DiskMass}

One of the most important works constraining the stellar masses of disk galaxies is the DiskMass survey \citep{bershady10,bershady11}.
DiskMass observes velocity dispersions across the disks of face-on galaxies.  It then assumes a disk thickness that is statistically consistent with
observations of edge-on galaxies \citep{Kregel} in order to estimate the stellar mass profile.  This method measures the vertical restoring
force to the disk, and is independent of the IMF.

The results of the DiskMass survey are summarized in the second and third rows.  
\citet{diskmass7} found a near constant NIR mass-to-light ratio of $\langle \ML^K \rangle = 0.31 \pm 0.07\;\MLsun$.
The small scatter and near-constancy of the mass-to-light ratio is consistent with our findings here.
The normalization differs by nearly a factor of two.

The rather sub-maximal disks found by the DiskMass survey imply that the dark matter halo contributes a non-negligible amount to
the observed velocity dispersions.  This will inevitably drive the implied mass of the disk down further.  \citet{diskmassSB} make a first estimate
of this effect, revising the mean mass-to-light ratio down to $\langle \ML^K \rangle = 0.24\;\MLsun$.  
This is a factor of $\sim 2.4$ times lower than what we find here.  The effect of the dark matter halos 
may be compensated somewhat by the presence of super-thin disks
in some galaxies \citep{Rook14}.  We will return to the factor of $\sim 2$ discrepancy
between the DiskMass and BTFR results after discussing other constraints.

\subsection{The Milky Way}
\label{sec:MW}

Our own Milky Way provides a unique environment in which we can hope to both count stars directly and constrain their mass from kinematics.
This is a rich field that is rapidly becoming richer; we consider here only a few relevant results.  \citet{flynn} use star counts from two independent
surveys to constrain Galactic structure.  Among other things, they estimate the mass-to-light ratio of the stellar disk of the Milky Way to be 
$\ML^V = 1.5 \pm 0.2$ and $\ML^I = 1.2 \pm 0.2\;\MLsun$.

\begin{figure*}
\epsscale{1.0}
\plotone{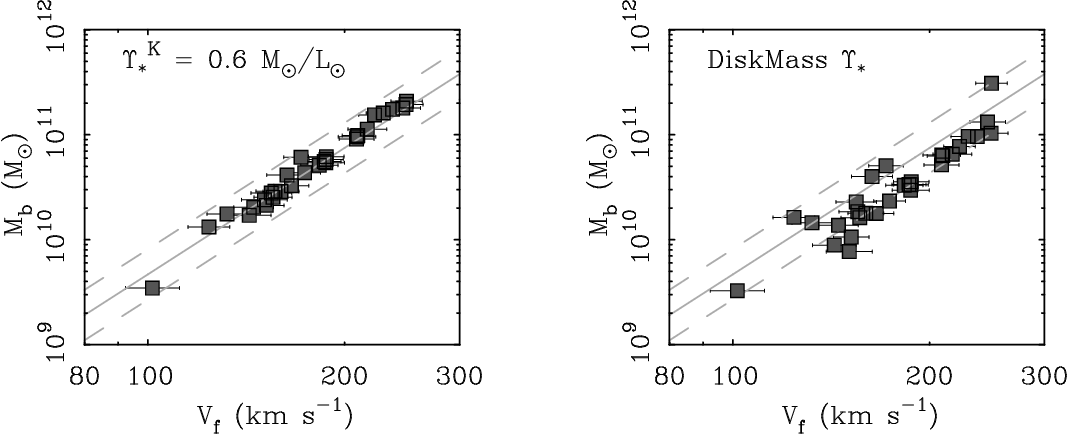}
\caption{The BTFR constructed from the galaxies in the DiskMass survey \citep{bershady10}.
Both panels are identical along the abscissa, with $V_f$ from \citet{diskmass6}.  They differ in stellar mass along the ordinate.  
The left panel shows the baryonic mass for stellar masses with a constant $K$-band mass-to-light ratio from population 
models \citep[$\ML^K = 0.6\;\MLsun$:][]{MS2014} plus the gas masses tabulated by \citet{diskmass7}.
The right panel uses the same gas masses, but adopts the stellar masses found by \citet{diskmass7}. 
The BTFR calibration of the gas rich galaxies \citep{M12} is shown as a solid line in both panels, with its $1 \sigma$
uncertainty as dashed lines.  
\label{DMBTFR}}
\end{figure*}

Starting from an initially exponential disk, \citet{M08} deformed the radial mass distribution of the Galactic disk in order to 
fit the terminal velocity curve in the fourth quadrant \citep{LBCM06,MGD07}.  
The result is a mass model with bumps and wiggles reminiscent of those observed in other spirals.
Indeed, the Centaurus arm is prominent in the inferred Milky Way mass profile.  Matching the details of the terminal velocity curve
requires the disk to be nearly maximal.  The stellar mass obtained in this way is $\sim 2\%$ greater than estimated by \citet{flynn}: 
$M_* = 5.48 \times 10^{10}\;\Msun$ for $R_0 = 8$ kpc.  To estimate the mass-to-light ratio, we adopt the total $I$-band luminosity estimated
by \citet{flynn}, $L_I = 4.5 \times10^{10}\;\Lsun$, and the $K$-band luminosity from \citet{DS01}, $L_K = 8.6 \times 10^{10}\;\Lsun$,
which includes a somewhat uncertain correction for the bulge fraction.  
The result is $\ML^I = 1.22\;\MLsun$ and $\ML^K = 0.63\;\MLsun$.

\citet{bovyrix} have measured the vertical restoring force to the disk $K_z(|Z|<1.1\;\mathrm{kpc})$ over a wide range of Galactocentric radii.
They obtain a mass for the stellar disk of $M_* = 4.6 \pm 0.3 \times 10^{10}\;\Msun$.  Combining this with the disk-only luminosity estimated 
by \citet{flynn}, $L_I^{disk} = 3.5 \times10^{10}\;\Lsun$, and $L_K^{disk} = 6.9 \times10^{10}\;\Lsun$ from \citet{DS01} 
yields $\ML^I = 1.31 \pm 0.09\;\MLsun$ and $\ML^K = 0.66 \pm 0.04\;\MLsun$.  
The bulge component is excluded from this calculation, as the measurement of \citet{bovyrix} is specific to the Galactic disk.  
The uncertainty in the mass-to-light ratio here is only that from the error in stellar mass quoted by \citet{bovyrix}, and does not include
the uncertainty in the luminosity.  The latter translates to roughly $\pm 0.2$ in \ML\ in absolute terms.

The three independent methods just discussed, star counts, the radial force in the disk, and the vertical force from the disk,
all find mass-to-light ratios for the Milky Way that are consistent within the errors.  They are also consistent with the result from the BTFR.  
They are not consistent with the low mass-to-light ratios implied by DiskMass, despite the similarity of the DiskMass approach to that of \citet{bovyrix}.

\subsection{The LMC}
\label{sec:LMC}

\citet{Eskew} perform star counts in the LMC with NIR data.  They sum up the mass of the stars observed to find $\ML^{[3.6]} = 0.5\;\MLsun$.
This method depends some on the IMF, as one does have to extrapolate to unseen, low mass stars.  As in the Milky Way, the stars are counted directly, albeit to a brighter limit, so the assumption on the IMF is less strong than in population synthesis models. \\

\subsection{Population Synthesis}
\label{sec:popsynth}

Stellar population synthesis is a well developed field.  It is possible to use our knowledge of stellar evolution to construct detailed models 
for the composite stellar populations of spiral galaxies for various assumed star formation histories and metallicities \citep[e.g.,][]{BZ03,PEGASE}.
For an assumed IMF, such models can be used to estimate the mass-to-light ratio of a population.  

One expects that as stars age, the population will redden and the mass-to-light ratio will increase.
Many models have been constructed to quantify this behavior.  A few relevant ones are noted in Table~\ref{MLresults}.

The models of \citet{Bell03}, \citet{Port04}, \citet{Zib09}, and \citet{IP13} have been explored in detail by \citet{MS2014}.
These all give comparable mass-to-light ratios in the optical portion of the spectrum: $1 < \ML^V < 1.5\;\MLsun$ evaluated at $B-V = 0.6$.  
Small differences there appear to be attributable largely to modest differences in the assumed IMF.  
However, the models have very different spectral energy distributions in the NIR, where agreement breaks down.  
In terms of \ML, \citet{Bell03} estimate $\ML^K = 0.73\;\MLsun$ for $B-V = 0.6$ while \citet{Zib09} give $\ML^K = 0.21\;\MLsun$.
This large discrepancy persists if we correct for difference in the IMF: scaling up the \citet{Zib09} estimate to match that of \citet{Bell03} in the
$V$-band has little impact in the NIR: $\ML^K = 0.21 \rightarrow 0.28\;\MLsun$.  This is still a factor of 2.6 lower than the $0.73\;\MLsun$ of \citet{Bell03}.
The models do not produce consistent spectral energy distributions \citep{Kriek10}.

\citet{MS2014} used multi-wavelength photometric data to test these various models.
Requiring self-consistency --- i.e., that the same mass of stars produce the observed luminosity in each band for every galaxy ---
leads to eq.~\ref{eqn:RevBell}.  This stipulation obliges all models to fall in the range $0.4 < \ML^{[3.6]} < 0.5\;\MLsun$.

\begin{figure*}
\epsscale{1.0}
\plotone{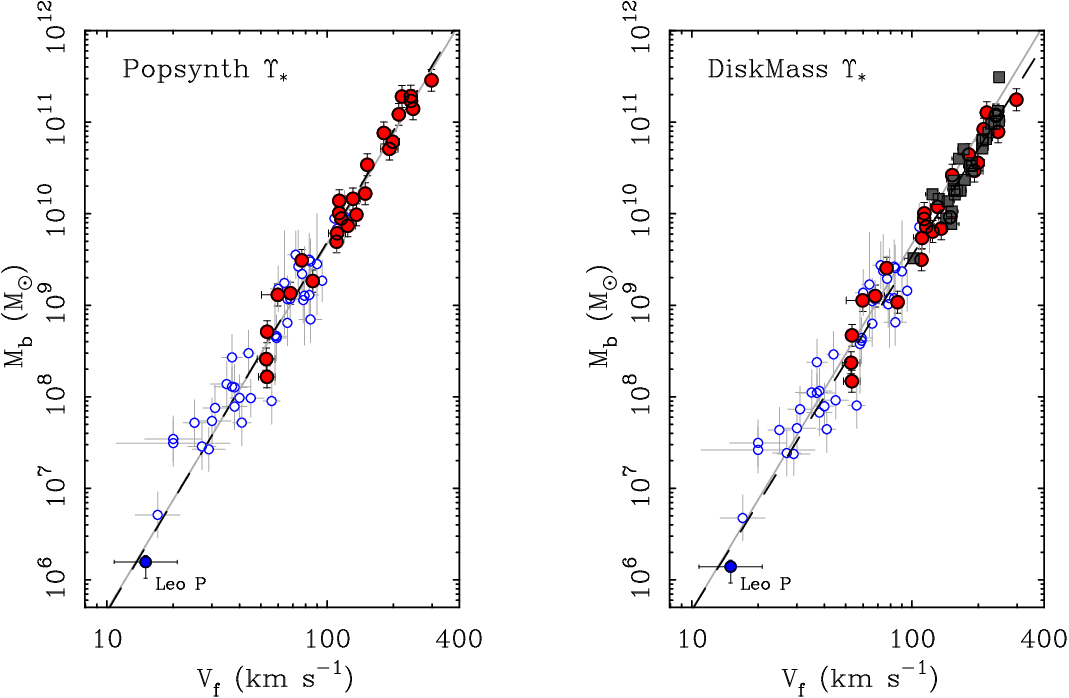}
\caption{The BTFR from the [3.6] sample (solid red circles) compared to the gas rich galaxy sample of \citet[open circles]{M12}
and the DiskMass data (squares).  The recently discovered, very low mass dwarf galaxy Leo P \citep{RLeoP,GLeoP} is also shown. 
In the left panel, the [3.6] data are shown with the population synthesis mass-to-light ratio $\ML^{[3.6]} = 0.47\;\MLsun$ and 0.12 dex uncertainty.
In the right panel, the DiskMass data are shown as in Fig.~\ref{DMBTFR}.  The stellar masses of the other two data sets are scaled down
to match the DiskMass scale ($\ML^{[3.6]} = 0.24\;\MLsun$ for the [3.6] data).  This makes little difference to the gas rich sample, but the
[3.6] sample shifts to match the DiskMass data.  The solid line in both panels is the fit to the gas rich data alone from \citet{M12}.
A fit to the combined gas rich plus [3.6] sample is shown as the dashed line in the left panel.
In the right panel, the dashed line is a fit to the gas rich plus DiskMass sample.  A tiny adjustment in slope (see Table~\ref{combinedfitstats})
over the six decades in mass illustrated here suffices to accommodate the DiskMass data with the gas rich data.
\label{6decades}}
\end{figure*}

The reasons for the difference between models are many, but the primary issue appears to be the treatment of 
TP-AGB stars.  The luminosity of this component is grossly exaggerated in the models of \citet{Zib09}.  The TP-AGB stars, as modeled,
produce a lot of NIR photons without contributing much in the optical nor representing much stellar mass.  
Consequently, the NIR mass-to-light ratios are underestimated.  A similar conclusion was reached by \citet{Zib13},
who sought but did not find the predicted spectral features of TP-AGB stars.
 
We constructed our own population models in \citet{SM14pop}.  These models have an empirically motivated distribution of stellar metallicities as
well as different ages corresponding to a variety of star formation histories.  The metallicity distribution weakens the dependence of the NIR
mass-to-light ratio on color relative to models built with a single metallicity.  Most of model parameter space is consistent 
with $\ML^{[3.6]} = 0.5\;\MLsun$.  Independently, \citet{Meidt2014} obtain $\ML^{[3.6]} = 0.6\;\MLsun$ albeit with a stronger color dependence.
These values are consistent with the majority of methods.  The exception to this is the DiskMass result, which we examine
further in the next section.

\subsection{The DiskMass BTFR}
\label{sec:dmbtfr}

Examination of Table~\ref{MLresults} suggests that nearly all methods are consistent with a NIR mass-to-light ratio 
$\ML^K \approx 0.57$ and $\ML^{[3.6]} \approx 0.45\;\MLsun$.  Since the population synthesis result for a Kroupa IMF
is consistent with the normalization independently required by the gas rich BTFR, it is tempting to conclude that the problem of stellar mass
in disk galaxies is essentially solved.  Unfortunately, the method of vertical velocity dispersions, as employed by the DiskMass project
\citep{bershady10,bershady11}, does not yield a consistent result.  

Since DiskMass provides a result that is independent of the IMF, it deserves more
consideration than the mutual consistency of the various population synthesis models. 
We can maintain that consistency while rescaling the IMF to match the DiskMass scale
with the translation $\Delta a = -0.29$.  So conceivably we simply need to adopt a different IMF
in the population modeling.  While this may be acceptable form a population synthesis perspective,
it is harder to reconcile with Milky Way constraints.  We check here whether we can reconcile
it with the BTFR.

We construct the BTFR from the DiskMass data \citep{diskmass6,diskmass7}
for two choices of mass-to-light ratio.  In one case, we use that measured by DiskMass \citep{diskmass7}. 
In the other, we adopt a constant mass-to-light ratio $\ML^K = 0.6\;\MLsun$ as suggested by population synthesis models \citep{MS2014}.
The result is shown in Fig.~\ref{DMBTFR}.
Unsurprisingly, the two BTFR are shifted with respect to each other due to the difference in the adopted mass-to-light ratios.

The BTFR constructed using the population synthesis value $\ML^K = 0.6\;\MLsun$ places the DiskMass galaxies directly
on the BTFR fit independently to gas rich galaxies by \citet{M12}.  This is not guaranteed to happen.  The gas rich galaxies are, 
by and large, much slower rotators than the bright spirals of the DiskMass sample.
The extrapolation of that fit will not work if the mass-to-light ratio is discordant.

Indeed, the data for the mass-to-light ratios determined by DiskMass are systematically shifted off the gas rich BTFR.
Most of the data parallel it, falling $\sim 1 \sigma$ below the fit line.  The DiskMass data are consistent with a slope 4 BTFR, 
but with a lower normalization.  One possibility is thus that the DiskMass \ML\ are too low due to some unknown systematic error,
and simply need to be shifted upwards to be consistent with other results.  

Looking closely at the DiskMass BTFR (right panel of Fig.~\ref{DMBTFR}), one may note that the data almost bifurcate into two parallel relations.
Most of the data fall low, but a half dozen galaxies stand to one side, consistent with the gas rich BTFR.
We do not know why the data might split like this, and suspect it may be indicative of a systematic error.
This bifurcation disappears when a higher, constant $K$-band mass-to-light ratio is adopted.
The scatter also decreases, but this is an inevitable result of setting the inclinations of the DiskMass galaxies
with the $K$-band LTFR in the first place \citep{diskmass7}.  As \ML\ grows, the gas becomes less important to $M_b$,
and one eventually just recovers a scaled version of the LTFR.  

So far, we have kept distinct the various samples: gas rich galaxies, the [3.6] sample of this paper, and the DiskMass data.
Another approach is to make a fit of the BTFR with the combined data.
The [3.6] and DiskMass data are consistent with each other for the same choice of mass-to-light ratio,
so we make two combined fits.  In one, we simply combine the [3.6] data as presented here with the gas rich data from \citet{M12}.
In the other, we combine the DiskMass data with the gas rich data, scaling the stellar masses of the gas rich galaxies down 
by a factor of two to be consistent with the DiskMass scale.
These fits are shown in Fig.~\ref{6decades} with fit parameters given in Table~\ref{combinedfitstats}.

\begin{deluxetable}{lcc}
\tablewidth{0pt}
\tablecaption{BTFR of Combined Samples}
\tablehead{
\colhead{Sample} & \colhead{$x$} & \colhead{$\log A$}  
}
\startdata
3.6 + gas rich & $4.04\pm0.09$ & $1.61\pm0.18$  \\
DiskMass + gas & $3.87\pm0.10$ & $1.80\pm0.21$ 
\enddata
\tablecomments{Fits of the form $\log M_b = \log A + x \log V_f$.
}
\label{combinedfitstats}
\end{deluxetable}

In combining the [3.6] and gas rich samples, we have been careful to exclude duplicate galaxies.
The gas rich galaxies from \citet{M12} are largely independent of the sample here, but there is some overlap.  
Of the 47 galaxies considered in \citet{M12} and the 26 in Table~\ref{basicdata}, there are 7 objects in common:
D631-7, DDO 168, IC 2574, F568-V1, F568-1, NGC 2403, and NGC 3198.
These seven are plotted (and fit) only once in Fig.~\ref{6decades}, as part of the [3.6] sample.

The combined fit obtained in this fashion is consistent with the previous fit to the gas rich data alone.
It is slightly steeper, as we have retained the stellar population estimate of $\ML^{[3.6]} = 0.47\;\MLsun$,
which is ever so slightly higher than the 0.45 of a straight extrapolation from the gas rich BTFR.
The fit lines are hardly distinguishable.

Indeed, the [3.6] data follow the extrapolation of the fit to the gas rich galaxies remarkably well. 
This happens with zero effort:  the location of the galaxies from Table \ref{basicdata} in the BTFR plane is determined entirely by the data and the
mass-to-light ratio provided by population modeling.  It is completely independent of the gas rich calibration. 

The gas rich BTFR calibration also extrapolates well downwards to lower masses.
This holds for both star dominated dwarf Spheroidals \citep{MWolf} and for the recently discovered gas rich dwarf Leo P \citep{GLeoP}.  
In Figure \ref{6decades} we adopt the distance of \citet{MQLeoP} when plotting the stellar \citep{RLeoP} and 
gas \citep{GLeoP} mass of Leo P so that the datum is independent of the fitted relation.  
Leo P falls along the extrapolation of the BTFR to lower rotation velocity just as the [3.6] data fall along the extrapolation to higher $V_f$.
The BTFR appears to be well described as a single power law over nearly six decades in mass.

In the combined DiskMass plus gas rich BTFR fit, the relation again appears to be well described as a single power law.
The chief difference is a slightly lower slope.  Instead of $x \approx 4.0$, we now find $x \approx 3.9$.

This change of slope is driven by the lower mass-to-light ratios favored by DiskMass.
The lower stellar masses adopted for the gas rich galaxies themselves make no perceptible difference.
The baryonic masses of these objects were already dominated by gas; for the new assumption this is slightly more true.
However, the gas rich galaxies are predominantly slow rotators, while the DiskMass sample is composed entirely of
fast rotators.  In order to fit the two data sets simultaneously, the BTFR simply pivots around the low mass end,
with the fitting routine selecting the slope necessary to hit the data at the high mass end.  The two lines illustrated
in the right panel of Fig.~\ref{6decades} are indistinguishable over the range constrained by the gas rich data,
but the slight difference in their slopes amounts to a factor of $\sim 2$ in \ML\ for the fast rotators.

It therefore appears that a factor of $\sim 2$ systematic uncertainty in stellar mass persists.
The natural conclusion of our work is that the population synthesis models are essentially correct for a reasonable (e.g., Kroupa) IMF.
However, we cannot exclude the lower mass scale indicated by the DiskMass data.  
The BTFR fit to DiskMass is within the uncertainties of the original gas rich BTFR fit.

The situation could be improved with better data.  Direct distance measurements of many gas rich galaxies would considerably
improve the calibration.  If the small intrinsic scatter of the BTFR persists, then the scatter will come down as the data improve.
The gas rich BTFR provides a promising new method to constrain the absolute stellar mass scale. 

\section{Conclusions}
\label{sec:conc}

We have weighed stellar disks in two distinct ways.
We have used a traditional population synthesis approach with updated color--mass-to-light ratio relations \citep[CMLR][]{MS2014}.
Separately, we have inferred the stellar mass required for bright galaxies to lie along the baryonic Tully-Fisher relation (BTFR) 
calibrated by gas rich galaxies.  These independent approaches yield consistent results.

We construct Tully-Fisher relations in the $V$, $I$, and [3.6] bands together with measurements of the outer, flat rotation velocity from
well resolved 21 cm data cubes.  Each band gives a distinct luminous Tully-Fisher relation (LTFR).  Each LTFR has a different slope,
intercept, and scatter.

The CMLR obtained by requiring photometric self-consistency by \citet{MS2014} successfully converts the different LTFR into a
consistent stellar mass Tully-Fisher relation (STFR).  This provides kinematic confirmation of the photometric results.
The slopes of the CMLR obtained by \citet{MS2014} must be approximately correct, or self-consistent STFR would not follow.

We use the scatter about the STFR to estimate the intrinsic scatter in the CMLR.
The scatter in stellar mass is 0.16 dex in $V$, 0.13 in $I$, and 0.12 in [3.6].  The trend for smaller scatter towards redder wavelengths
follows the expectation of stellar population models.  Indeed, the modest amount of scatter in the near-infrared (NIR) is consistent with the models 
of \citet{BdJ01} and \citet{Port04}, who estimate 0.1 and 0.15 dex in the $K$-band, respectively.  This consistency implies that the variation
in star formation history from galaxy to galaxy suffices to explain the observed scatter, leaving little room for other sources of scatter.
This in turn implies that the IMF is effectively universal, at least when averaged over entire galaxies.

The stellar masses estimated with our CMLR also lead to nice convergence in the BTFR.
The BTFR obtained in this fashion is consistent with that obtained independently from gas rich galaxies \citep{M12}.
The slope is steep, being very close to $x = 4$ in $M_b \propto V_f^x$.  
There is no indication of a bend that transitions between a steeper slope at low mass and a shallower slope at high mass as 
suggested by some models \citep[e.g.,][]{TGKPR}.

The intrinsic scatter of the BTFR is very small, consistent with zero.
Effectively all of the observed scatter can be attributed to observational uncertainties and the scatter in stellar mass-to-light ratios.
Ignoring these, the upper limit on intrinsic scatter is that observed at [3.6]: 0.13 dex.
This is difficult to understand in the context of dark matter models, in which there should be a considerable amount of
scatter from halo to halo \citep{EL96,MdB98a,BullockTF}.

The BTFR calibrated by gas dominated galaxies provides an independent constraint on stellar mass.  
This approach also yields a typical $3.6\micron$ mass-to-light ratio of $\ML^{[3.6]} = 0.45\;\MLsun$,
with a corresponding value in the $K_s$-band of $\ML^K = 0.57\;\MLsun$.
The scatter about these typical values is 0.12 dex, consistent with the anticipated variation in star formation 
histories \citep{BdJ01,Port04,KelsonCLT}.

An important consequence of this result is that the NIR luminosity of a galaxy is a good proxy for its stellar mass.
Workers wishing to estimate the the stellar mass of galaxies would do well to measure this quantity and 
adopt a constant NIR mass-to-light ratio as calibrated here.
If a NIR luminosity is not available, an optical luminosity is also a good proxy, albeit with more scatter and some color dependence.
Contrary to our initial hope, there appears to be little value added in fitting the entire spectral energy distribution of galaxies,
at least so far as their mass-to-light ratio is concerned: 
most of the information about the mass-to-light ratio is carried by a single color like $B-V$.  Further color-based corrections are
more likely to reveal systematic uncertainties in stellar population models than they are to improve the estimate of stellar mass,
at least at this juncture.

The BTFR provides a new method to calibrate the absolute scale of stellar mass in rotating galaxies.
It provides a useful constraint that is consistent with stellar population synthesis models.
The accuracy of this method could be considerably improved with straightforward observations,
such as direct distance measurements of gas rich galaxies.

\acknowledgements  
We thank the referee, Dennis Zaritsky, for a thoughtful review that led us to broaden the scope of this paper.
We also thank Matt Bershady and Rob Swaters for conversations about the DiskMass project.
This work is based in part on observations made with the Spitzer Space Telescope, which is operated by the Jet Propulsion Laboratory, 
California Institute of Technology under a contract with NASA. Support for this work was provided by NASA through an award issued by 
JPL/Caltech. Other aspects of this work were supported in part by NASA ADAP grant NNX11AF89G.  
This research has made use of the NASA/IPAC Extragalactic Database (NED) which is operated by the Jet Propulsion Laboratory, 
California Institute of Technology, under contract with the National Aeronautics and Space Administration.

\bibliography{stms}

\begin{thebibliography}{76}
\expandafter\ifx\csname natexlab\endcsname\relax\def\natexlab#1{#1}\fi

\bibitem[{{Begum} {et~al.}(2008){Begum}, {Chengalur}, {Karachentsev}, \&
  {Sharina}}]{begum}
{Begum}, A., {Chengalur}, J.~N., {Karachentsev}, I.~D., \& {Sharina}, M.~E.
  2008, \mnras, 386, 138

\bibitem[{{Bell} \& {de Jong}(2001)}]{BdJ01}
{Bell}, E.~F., \& {de Jong}, R.~S. 2001, \apj, 550, 212

\bibitem[{{Bell} {et~al.}(2003){Bell}, {McIntosh}, {Katz}, \&
  {Weinberg}}]{Bell03}
{Bell}, E.~F., {McIntosh}, D.~H., {Katz}, N., \& {Weinberg}, M.~D. 2003, \apjs,
  149, 289

\bibitem[{{Bershady} {et~al.}(2011){Bershady}, {Martinsson}, {Verheijen},
  {Westfall}, {Andersen}, \& {Swaters}}]{bershady11}
{Bershady}, M.~A., {Martinsson}, T.~P.~K., {Verheijen}, M.~A.~W., {Westfall},
  K.~B., {Andersen}, D.~R., \& {Swaters}, R.~A. 2011, \apjl, 739, L47

\bibitem[{{Bershady} {et~al.}(2010){Bershady}, {Verheijen}, {Swaters},
  {Andersen}, {Westfall}, \& {Martinsson}}]{bershady10}
{Bershady}, M.~A., {Verheijen}, M.~A.~W., {Swaters}, R.~A., {Andersen}, D.~R.,
  {Westfall}, K.~B., \& {Martinsson}, T. 2010, \apj, 716, 198

\bibitem[{{Bottema} {et~al.}(2002){Bottema}, {Pesta{\~n}a}, {Rothberg}, \&
  {Sanders}}]{bottema02}
{Bottema}, R., {Pesta{\~n}a}, J.~L.~G., {Rothberg}, B., \& {Sanders}, R.~H.
  2002, \aap, 393, 453

\bibitem[{{Bovy} \& {Rix}(2013)}]{bovyrix}
{Bovy}, J., \& {Rix}, H.-W. 2013, \apj, 779, 115

\bibitem[{{Bruzual} \& {Charlot}(2003)}]{BZ03}
{Bruzual}, G., \& {Charlot}, S. 2003, \mnras, 344, 1000

\bibitem[{{Bullock} {et~al.}(2001){Bullock}, {Kolatt}, {Sigad}, {Somerville},
  {Kravtsov}, {Klypin}, {Primack}, \& {Dekel}}]{BullockTF}
{Bullock}, J.~S., {Kolatt}, T.~S., {Sigad}, Y., {Somerville}, R.~S.,
  {Kravtsov}, A.~V., {Klypin}, A.~A., {Primack}, J.~R., \& {Dekel}, A. 2001,
  \mnras, 321, 559

\bibitem[{{Chabrier}(2003)}]{chabrier}
{Chabrier}, G. 2003, \apjl, 586, L133

\bibitem[{{Conroy} \& {Gunn}(2010)}]{conroy}
{Conroy}, C., \& {Gunn}, J.~E. 2010, \apj, 712, 833

\bibitem[{{Dale} {et~al.}(2005){Dale}, {Bendo}, {Engelbracht}, {Gordon},
  {Regan}, {Armus}, {Cannon}, {Calzetti}, {Draine}, {Helou}, {Joseph},
  {Kennicutt}, {Li}, {Murphy}, {Roussel}, {Walter}, {Hanson}, {Hollenbach},
  {Jarrett}, {Kewley}, {Lamanna}, {Leitherer}, {Meyer}, {Rieke}, {Rieke},
  {Sheth}, {Smith}, \& {Thornley}}]{DSINGS}
{Dale}, D.~A., {et~al.} 2005, \apj, 633, 857

\bibitem[{{de Blok} \& {McGaugh}(1996)}]{dBM1996}
{de Blok}, W.~J.~G., \& {McGaugh}, S.~S. 1996, \apjl, 469, L89

\bibitem[{{de Blok} {et~al.}(1996){de Blok}, {McGaugh}, \& {van der
  Hulst}}]{dBMH96}
{de Blok}, W.~J.~G., {McGaugh}, S.~S., \& {van der Hulst}, J.~M. 1996, \mnras,
  283, 18

\bibitem[{{de Blok} {et~al.}(2008){de Blok}, {Walter}, {Brinks},
  {Trachternach}, {Oh}, \& {Kennicutt}}]{THINGS}
{de Blok}, W.~J.~G., {Walter}, F., {Brinks}, E., {Trachternach}, C., {Oh},
  S.-H., \& {Kennicutt}, R.~C. 2008, \aj, 136, 2648

\bibitem[{{Drimmel} \& {Spergel}(2001)}]{DS01}
{Drimmel}, R., \& {Spergel}, D.~N. 2001, \apj, 556, 181

\bibitem[{{Dutton}(2012)}]{Dutton2012}
{Dutton}, A.~A. 2012, \mnras, 424, 3123

\bibitem[{{Eisenstein} \& {Loeb}(1996)}]{EL96}
{Eisenstein}, D.~J., \& {Loeb}, A. 1996, \apj, 459, 432

\bibitem[{{Eskew} {et~al.}(2012){Eskew}, {Zaritsky}, \& {Meidt}}]{Eskew}
{Eskew}, M., {Zaritsky}, D., \& {Meidt}, S. 2012, \aj, 143, 139

\bibitem[{{Flynn} {et~al.}(2006){Flynn}, {Holmberg}, {Portinari}, {Fuchs}, \&
  {Jahrei{\ss}}}]{flynn}
{Flynn}, C., {Holmberg}, J., {Portinari}, L., {Fuchs}, B., \& {Jahrei{\ss}}, H.
  2006, \mnras, 372, 1149

\bibitem[{{Giovanelli} {et~al.}(2013){Giovanelli}, {Haynes}, {Adams}, {Cannon},
  {Rhode}, {Salzer}, {Skillman}, {Bernstein-Cooper}, \& {McQuinn}}]{GLeoP}
{Giovanelli}, R., {et~al.} 2013, \aj, 146, 15

\bibitem[{{Governato} {et~al.}(2007){Governato}, {Willman}, {Mayer}, {Brooks},
  {Stinson}, {Valenzuela}, {Wadsley}, \& {Quinn}}]{fabioTF}
{Governato}, F., {Willman}, B., {Mayer}, L., {Brooks}, A., {Stinson}, G.,
  {Valenzuela}, O., {Wadsley}, J., \& {Quinn}, T. 2007, \mnras, 374, 1479

\bibitem[{{Into} \& {Portinari}(2013)}]{IP13}
{Into}, T., \& {Portinari}, L. 2013, \mnras, 430, 2715

\bibitem[{{Kelson}(2014)}]{KelsonCLT}
{Kelson}, D.~D. 2014, arXiv:1406.5191

\bibitem[{{Kennicutt} {et~al.}(2003){Kennicutt}, {Armus}, {Bendo}, {Calzetti},
  {Dale}, {Draine}, {Engelbracht}, {Gordon}, {Grauer}, {Helou}, {Hollenbach},
  {Jarrett}, {Kewley}, {Leitherer}, {Li}, {Malhotra}, {Regan}, {Rieke},
  {Rieke}, {Roussel}, {Smith}, {Thornley}, \& {Walter}}]{KSINGS}
{Kennicutt}, Jr., R.~C., {et~al.} 2003, \pasp, 115, 928

\bibitem[{{Kregel} {et~al.}(2005){Kregel}, {van der Kruit}, \&
  {Freeman}}]{Kregel}
{Kregel}, M., {van der Kruit}, P.~C., \& {Freeman}, K.~C. 2005, \mnras, 358,
  503

\bibitem[{{Kriek} {et~al.}(2010){Kriek}, {Labb{\'e}}, {Conroy}, {Whitaker},
  {van Dokkum}, {Brammer}, {Franx}, {Illingworth}, {Marchesini}, {Muzzin},
  {Quadri}, \& {Rudnick}}]{Kriek10}
{Kriek}, M., {et~al.} 2010, \apjl, 722, L64

\bibitem[{{Kroupa}(1998)}]{kroupa}
{Kroupa}, P. 1998, in Astronomical Society of the Pacific Conference Series,
  Vol. 134, Brown Dwarfs and Extrasolar Planets, ed. R.~{Rebolo}, E.~L.
  {Martin}, \& M.~R. {Zapatero Osorio}, 483

\bibitem[{{Le Borgne} {et~al.}(2004){Le Borgne}, {Rocca-Volmerange},
  {Prugniel}, {Lan{\c c}on}, {Fioc}, \& {Soubiran}}]{PEGASE}
{Le Borgne}, D., {Rocca-Volmerange}, B., {Prugniel}, P., {Lan{\c c}on}, A.,
  {Fioc}, M., \& {Soubiran}, C. 2004, \aap, 425, 881

\bibitem[{{Leroy} {et~al.}(2008){Leroy}, {Walter}, {Brinks}, {Bigiel}, {de
  Blok}, {Madore}, \& {Thornley}}]{leroy}
{Leroy}, A.~K., {Walter}, F., {Brinks}, E., {Bigiel}, F., {de Blok}, W.~J.~G.,
  {Madore}, B., \& {Thornley}, M.~D. 2008, \aj, 136, 2782

\bibitem[{{Luna} {et~al.}(2006){Luna}, {Bronfman}, {Carrasco}, \&
  {May}}]{LBCM06}
{Luna}, A., {Bronfman}, L., {Carrasco}, L., \& {May}, J. 2006, \apj, 641, 938

\bibitem[{{Maraston}(2005)}]{maraston05}
{Maraston}, C. 2005, \mnras, 362, 799

\bibitem[{{Marigo} {et~al.}(2008){Marigo}, {Girardi}, {Bressan}, {Groenewegen},
  {Silva}, \& {Granato}}]{Marigo2008}
{Marigo}, P., {Girardi}, L., {Bressan}, A., {Groenewegen}, M.~A.~T., {Silva},
  L., \& {Granato}, G.~L. 2008, \aap, 482, 883

\bibitem[{{Martinsson} {et~al.}(2013{\natexlab{a}}){Martinsson}, {Verheijen},
  {Westfall}, {Bershady}, {Andersen}, \& {Swaters}}]{diskmass7}
{Martinsson}, T.~P.~K., {Verheijen}, M.~A.~W., {Westfall}, K.~B., {Bershady},
  M.~A., {Andersen}, D.~R., \& {Swaters}, R.~A. 2013{\natexlab{a}}, \aap, 557,
  A131

\bibitem[{{Martinsson} {et~al.}(2013{\natexlab{b}}){Martinsson}, {Verheijen},
  {Westfall}, {Bershady}, {Schechtman-Rook}, {Andersen}, \&
  {Swaters}}]{diskmass6}
{Martinsson}, T.~P.~K., {Verheijen}, M.~A.~W., {Westfall}, K.~B., {Bershady},
  M.~A., {Schechtman-Rook}, A., {Andersen}, D.~R., \& {Swaters}, R.~A.
  2013{\natexlab{b}}, \aap, 557, A130

\bibitem[{{Matthews} {et~al.}(1998){Matthews}, {van Driel}, \&
  {Gallagher}}]{MvDG98}
{Matthews}, L.~D., {van Driel}, W., \& {Gallagher}, III, J.~S. 1998, \aj, 116,
  2196

\bibitem[{{McClure-Griffiths} \& {Dickey}(2007)}]{MGD07}
{McClure-Griffiths}, N.~M., \& {Dickey}, J.~M. 2007, \apj, 671, 427

\bibitem[{{McGaugh}(2005)}]{M05}
{McGaugh}, S.~S. 2005, \apj, 632, 859

\bibitem[{{McGaugh}(2008)}]{M08}
---. 2008, \apj, 683, 137

\bibitem[{{McGaugh}(2011)}]{M11}
---. 2011, Phys.~Rev.~Lett., 106, 121303

\bibitem[{{McGaugh}(2012)}]{M12}
---. 2012, \aj, 143, 40

\bibitem[{{McGaugh} \& {de Blok}(1998)}]{MdB98a}
{McGaugh}, S.~S., \& {de Blok}, W.~J.~G. 1998, \apj, 499, 41

\bibitem[{{McGaugh} \& {Schombert}(2014)}]{MS2014}
{McGaugh}, S.~S., \& {Schombert}, J.~M. 2014, \aj, 148, 77

\bibitem[{{McGaugh} {et~al.}(2000){McGaugh}, {Schombert}, {Bothun}, \& {de
  Blok}}]{btforig}
{McGaugh}, S.~S., {Schombert}, J.~M., {Bothun}, G.~D., \& {de Blok}, W.~J.~G.
  2000, \apjl, 533, L99

\bibitem[{{McGaugh} \& {Wolf}(2010)}]{MWolf}
{McGaugh}, S.~S., \& {Wolf}, J. 2010, \apj, 722, 248

\bibitem[{{McQuinn} {et~al.}(2013){McQuinn}, {Skillman}, {Berg}, {Cannon},
  {Salzer}, {Adams}, {Dolphin}, {Giovanelli}, {Haynes}, \& {Rhode}}]{MQLeoP}
{McQuinn}, K.~B.~W., {et~al.} 2013, \aj, 146, 145

\bibitem[{{Meidt} {et~al.}(2014){Meidt}, {Schinnerer}, {van de Ven},
  {Zaritsky}, {Peletier}, {Knapen}, {Sheth}, {Regan}, {Querejeta},
  {Mu{\~n}oz-Mateos}, {Kim}, {Hinz}, {Gil de Paz}, {Athanassoula}, {Bosma},
  {Buta}, {Cisternas}, {Ho}, {Holwerda}, {Skibba}, {Laurikainen}, {Salo},
  {Gadotti}, {Laine}, {Erroz-Ferrer}, {Comer{\'o}n}, {Men{\'e}ndez-Delmestre},
  {Seibert}, \& {Mizusawa}}]{Meidt2014}
{Meidt}, S.~E., {et~al.} 2014, \apj, 788, 144

\bibitem[{{Melbourne} {et~al.}(2012){Melbourne}, {Williams}, {Dalcanton},
  {Rosenfield}, {Girardi}, {Marigo}, {Weisz}, {Dolphin}, {Boyer}, {Olsen},
  {Skillman}, \& {Seth}}]{Melbourne12}
{Melbourne}, J., {et~al.} 2012, \apj, 748, 47

\bibitem[{{Milgrom}(1983)}]{milgrom83}
{Milgrom}, M. 1983, {\apj}, 270, 371

\bibitem[{{Noordermeer} \& {Verheijen}(2007)}]{noordTF}
{Noordermeer}, E., \& {Verheijen}, M.~A.~W. 2007, \mnras, 381, 1463

\bibitem[{{Oh} {et~al.}(2008){Oh}, {de Blok}, {Walter}, {Brinks}, \&
  {Kennicutt}}]{OhThings}
{Oh}, S., {de Blok}, W.~J.~G., {Walter}, F., {Brinks}, E., \& {Kennicutt},
  R.~C. 2008, \aj, 136, 2761

\bibitem[{{Pfenniger} \& {Revaz}(2005)}]{pfennBTF}
{Pfenniger}, D., \& {Revaz}, Y. 2005, \aap, 431, 511

\bibitem[{{Pizagno} {et~al.}(2007){Pizagno}, {Prada}, {Weinberg}, {Rix},
  {Pogge}, {Grebel}, {Harbeck}, {Blanton}, {Brinkmann}, \& {Gunn}}]{Pizagno07}
{Pizagno}, J., {et~al.} 2007, \aj, 134, 945

\bibitem[{{Portinari} {et~al.}(2004){Portinari}, {Sommer-Larsen}, \&
  {Tantalo}}]{Port04}
{Portinari}, L., {Sommer-Larsen}, J., \& {Tantalo}, R. 2004, \mnras, 347, 691

\bibitem[{{Rhode} {et~al.}(2013){Rhode}, {Salzer}, {Haurberg}, {Van Sistine},
  {Young}, {Haynes}, {Giovanelli}, {Cannon}, {Skillman}, {McQuinn}, \&
  {Adams}}]{RLeoP}
{Rhode}, K.~L., {et~al.} 2013, \aj, 145, 149

\bibitem[{{Schechtman-Rook} \& {Bershady}(2014)}]{Rook14}
{Schechtman-Rook}, A., \& {Bershady}, M.~A. 2014, \apj, 795, 136

\bibitem[{{Schombert} {et~al.}(2011){Schombert}, {Maciel}, \&
  {McGaugh}}]{SMM11}
{Schombert}, J., {Maciel}, T., \& {McGaugh}, S. 2011, Advances in Astronomy,
  2011

\bibitem[{{Schombert} \& {McGaugh}(2014{\natexlab{a}})}]{SM14pop}
{Schombert}, J., \& {McGaugh}, S. 2014{\natexlab{a}}, PASA, 31, 36

\bibitem[{{Schombert} \& {McGaugh}(2014{\natexlab{b}})}]{SM2014}
---. 2014{\natexlab{b}}, PASA, 31, 11

\bibitem[{{Schombert} \& {Rakos}(2009)}]{rakosschombert09a}
{Schombert}, J., \& {Rakos}, K. 2009, \aj, 137, 528

\bibitem[{{Sorce} {et~al.}(2012){Sorce}, {Tully}, \& {Courtois}}]{Sorce2012}
{Sorce}, J.~G., {Tully}, R.~B., \& {Courtois}, H.~M. 2012, \apjl, 758, L12

\bibitem[{{Stark} {et~al.}(2009){Stark}, {McGaugh}, \& {Swaters}}]{stark}
{Stark}, D.~V., {McGaugh}, S.~S., \& {Swaters}, R.~A. 2009, \aj, 138, 392

\bibitem[{{Swaters} {et~al.}(2014){Swaters}, {Bershady}, {Martinsson},
  {Westfall}, {Andersen}, \& {Verheijen}}]{diskmassSB}
{Swaters}, R.~A., {Bershady}, M.~A., {Martinsson}, T.~P.~K., {Westfall}, K.~B.,
  {Andersen}, D.~R., \& {Verheijen}, M.~A.~W. 2014, \apjl, 797, L28

\bibitem[{{Swaters} {et~al.}(2009){Swaters}, {Sancisi}, {van Albada}, \& {van
  der Hulst}}]{swaterswhisp}
{Swaters}, R.~A., {Sancisi}, R., {van Albada}, T.~S., \& {van der Hulst}, J.~M.
  2009, \aap, 493, 871

\bibitem[{{Trachternach} {et~al.}(2009){Trachternach}, {de Blok}, {McGaugh},
  {van der Hulst}, \& {Dettmar}}]{trach}
{Trachternach}, C., {de Blok}, W.~J.~G., {McGaugh}, S.~S., {van der Hulst},
  J.~M., \& {Dettmar}, R. 2009, \aap, 505, 577

\bibitem[{{Trujillo-Gomez} {et~al.}(2011){Trujillo-Gomez}, {Klypin}, {Primack},
  \& {Romanowsky}}]{TGKPR}
{Trujillo-Gomez}, S., {Klypin}, A., {Primack}, J., \& {Romanowsky}, A.~J. 2011,
  \apj, 742, 16

\bibitem[{{Tully} \& {Fisher}(1977)}]{TForig}
{Tully}, R.~B., \& {Fisher}, J.~R. 1977, \aap, 54, 661

\bibitem[{{Tully} {et~al.}(1998){Tully}, {Pierce}, {Huang}, {Saunders},
  {Verheijen}, \& {Witchalls}}]{TFdust}
{Tully}, R.~B., {Pierce}, M.~J., {Huang}, J., {Saunders}, W., {Verheijen},
  M.~A.~W., \& {Witchalls}, P.~L. 1998, \aj, 115, 2264

\bibitem[{{Tully} {et~al.}(2009){Tully}, {Rizzi}, {Shaya}, {Courtois},
  {Makarov}, \& {Jacobs}}]{EDD}
{Tully}, R.~B., {Rizzi}, L., {Shaya}, E.~J., {Courtois}, H.~M., {Makarov},
  D.~I., \& {Jacobs}, B.~A. 2009, \aj, 138, 323

\bibitem[{{Verheijen} \& {de Blok}(1999)}]{VdB1999}
{Verheijen}, M., \& {de Blok}, E. 1999, \apss, 269, 673

\bibitem[{{Verheijen}(2001)}]{verhTF}
{Verheijen}, M.~A.~W. 2001, \apj, 563, 694

\bibitem[{{Walter} {et~al.}(2008){Walter}, {Brinks}, {de Blok}, {Bigiel},
  {Kennicutt}, {Thornley}, \& {Leroy}}]{FTHINGS}
{Walter}, F., {Brinks}, E., {de Blok}, W.~J.~G., {Bigiel}, F., {Kennicutt},
  R.~C., {Thornley}, M.~D., \& {Leroy}, A. 2008, \aj, 136, 2563

\bibitem[{{Weiner} {et~al.}(2006){Weiner}, {Willmer}, {Faber}, {Harker},
  {Kassin}, {Phillips}, {Melbourne}, {Metevier}, {Vogt}, \& {Koo}}]{benfit}
{Weiner}, B.~J., {et~al.} 2006, \apj, 653, 1049

\bibitem[{{Zaritsky} {et~al.}(2014){Zaritsky}, {Courtois}, {Mu{\~n}oz-Mateos},
  {Sorce}, {Erroz-Ferrer}, {Comer{\'o}n}, {Gadotti}, {Gil de Paz}, {Hinz},
  {Laurikainen}, {Kim}, {Laine}, {Men{\'e}ndez-Delmestre}, {Mizusawa}, {Regan},
  {Salo}, {Seibert}, {Sheth}, {Athanassoula}, {Bosma}, {Cisternas}, {Ho}, \&
  {Holwerda}}]{Zarit2014}
{Zaritsky}, D., {et~al.} 2014, \aj, 147, 134

\bibitem[{{Zibetti} {et~al.}(2009){Zibetti}, {Charlot}, \& {Rix}}]{Zib09}
{Zibetti}, S., {Charlot}, S., \& {Rix}, H.-W. 2009, \mnras, 400, 1181

\bibitem[{{Zibetti} {et~al.}(2013){Zibetti}, {Gallazzi}, {Charlot}, {Pierini},
  \& {Pasquali}}]{Zib13}
{Zibetti}, S., {Gallazzi}, A., {Charlot}, S., {Pierini}, D., \& {Pasquali}, A.
  2013, \mnras, 428, 1479

\end{thebibliography}

\bibliographystyle{apj}

\end{document}